\documentclass{IEEEoj}
\usepackage{cite}
\usepackage{amsmath,amssymb,amsfonts}
\usepackage{algorithmic}
\usepackage{graphicx,color}
\usepackage{textcomp}
\usepackage{url}
\usepackage{subcaption}
\def\BibTeX{{\rm B\kern-.05em{\sc i\kern-.025em b}\kern-.08em
    T\kern-.1667em\lower.7ex\hbox{E}\kern-.125emX}}
\AtBeginDocument{\definecolor{ojcolor}{cmyk}{0.93,0.59,0.15,0.02}}
\def\OJlogo{\vspace{-4pt}\hskip-4pt\includegraphics[height=18pt]{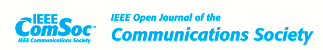}}
\begin{document}
\receiveddate{XX Month, XXXX}
\reviseddate{XX Month, XXXX}
\accepteddate{XX Month, XXXX}
\publisheddate{XX Month, XXXX}
\currentdate{XX Month, XXXX}
\doiinfo{OJCOMS.2026.XXXXXX}

\title{Mobile Base Station Positioning in Smart Ports Based on Kriged Sparse Measurements and Obstacle Inference}

\author{PAULO FURTADO CORREIA\IEEEauthorrefmark{1}, ANDRÉ COELHO\IEEEauthorrefmark{1}\IEEEmembership{(Member, IEEE)}, and MANUEL RICARDO\IEEEauthorrefmark{1}
\IEEEmembership{(Member, IEEE)}}
\affil{INESC TEC, Faculdade de Engenharia da Universidade do Porto, Portugal}
\corresp{CORRESPONDING AUTHOR: Paulo Furtado Correia (e-mail: paulo.j.correia@inesctec.pt).}
\authornote{This work is co-financed by Component 5 – Capitalization and Business Innovation integrated in the Resilience Dimension of the Recovery
and Resilience Plan within the scope of the Recovery and Resilience Mechanism (MRR) of the European Union (EU), framed in the Next
Generation EU, for the period 2021 – 2026, within project NEXUS, with reference 53.}
\markboth{Mobile Base Station Positioning in Smart Ports Based on Kriged Sparse Measurements and Obstacle Inference}{P. F. Correia \emph{et al.}}

\begin{abstract}
Smart-port wireless networks suffer from severe and dynamic radio blockage caused by container stacks and industrial structures, making efficient mobile integrated access and backhaul (MIAB) deployment challenging without accurate environmental awareness. Existing approaches typically rely on prior obstacle maps, explicit geometry information, or computationally intensive propagation models that limit adaptability in operational deployments.
This paper presents \textsc{DOCKING}, a radio environment map (REM)-driven framework that converts sparse radio measurements into optimization-ready obstacle representations for MIAB deployment. The proposed approach infers propagation-relevant obstacle abstractions directly from reconstructed REMs, avoiding the need for obstacle-geometry databases while relying on known network parameters and sparse radio measurements. Sparse reference signal received power (RSRP) and signal-to-interference-plus-noise ratio (SINR) observations are reconstructed through Ordinary Kriging (OKG), after which dominant attenuation regions are identified and approximated by compact cuboidal blockage characterization.
The inferred geometry is incorporated into a backhaul-aware optimization stage that jointly determines MIAB placement, user-equipment (UE) association, and backhaul selection. Under realistic smart-port conditions, REM reconstruction achieves prediction errors below $3\,\mathrm{dB}$ at the $90$th percentile using only $15\%$ spatial sampling, while obstacle characterization exceeds $85\%$ true-positive coverage. Capacity gains reach up to $150\%$ in sparse deployment scenarios, and a fast Genetic Algorithm converges within $5$--$15\,\mathrm{s}$ per network snapshot. A proof-of-concept field campaign further corroborates the proposed REM-to-obstacle-to-MIAB workflow using real measurements, showing throughput trends consistent with the optimization predictions.
These results demonstrate that sparse radio measurements can provide sufficient environmental awareness to support practical obstacle-aware MIAB deployment in obstruction-prone industrial environments.
\end{abstract}
\begin{IEEEkeywords}
Integrated access and backhaul (IAB),
mobile integrated access and backhaul (MIAB),
radio environment maps (REM),
ordinary kriging,
obstacle characterization,
smart ports,
genetic algorithms
\end{IEEEkeywords}
\maketitle

\section{INTRODUCTION}
\IEEEPARstart{T}{he} transition toward sixth-generation (6G) wireless systems is driven by the need for reliable, high-capacity, and adaptive connectivity in dynamic environments~\cite{b1}. Among these, smart seaports are undergoing rapid digital transformation through industrial Internet of Things (IIoT) sensing, autonomous guided vehicles (AGVs), remote crane systems, real-time asset tracking, and dense video surveillance~\cite{b2,b3}. These services place sustained pressure on radio resources and make network planning a critical challenge.
\begin{figure*}[t]
    \centering
    \includegraphics[width=\textwidth]{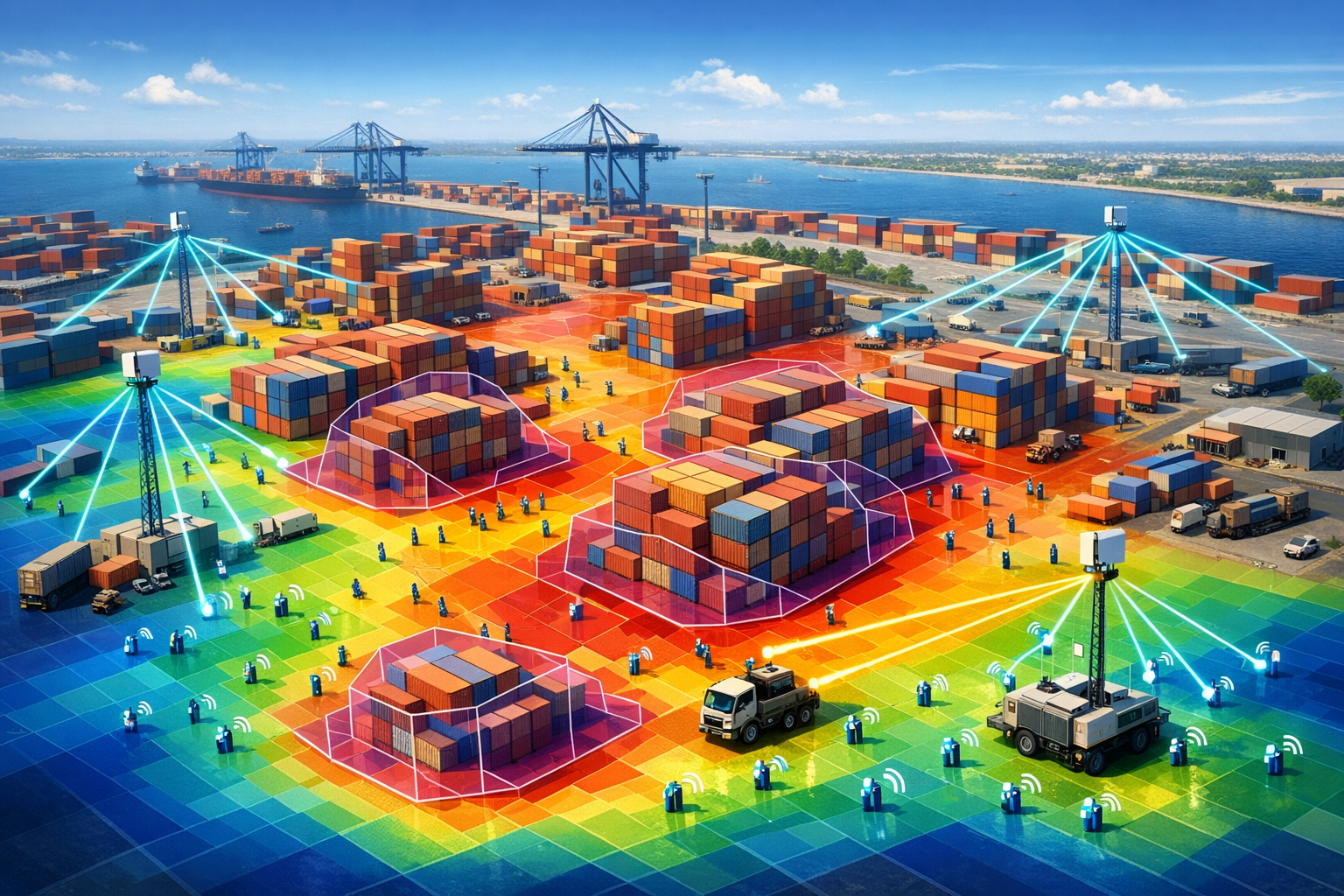}
    \caption{Panoramic view of a smart port, showing its dynamic logistics landscape. FIABs provide wireless access to UEs across the terminal, while a MIAB mounted on a terrestrial vehicle extends coverage. High-attenuation regions can be detected through \textsc{DOCKING}.}
    \label{fig:environment}
\end{figure*}

Despite the comparatively favorable propagation characteristics of sub-6~GHz
bands, seaport environments present substantial challenges arising from their
physical structure. Dense stacks of metallic shipping containers---often
arranged four to six units high in irregular formations across open yard
areas---act as dominant electromagnetic obstacles, producing severe shadowing,
diffraction, and multipath effects that introduce deep signal voids across
large portions of the terminal~\cite{b4}. An illustrative overview of the
environment is shown in Fig.~\ref{fig:environment}.
These obstacles are dynamic: loading, unloading, and repositioning operations continuously alter the radio environment, making exhaustive drive-test planning and full-area RSRP measurement impractical at operational scale~\cite{b5}.
Consequently, existing approaches do not jointly address radio map reconstruction, obstacle geometry abstraction, and mobile relay deployment from a single sparse measurement set in the absence of prior obstacle-geometry information. This gap is the focus of this paper. 

By reducing the dependence on precomputed maps and dense site surveys, the network can dynamically infer propagation-relevant shadowing structures from their radio-frequency footprints. To address this challenge within our pipeline, geostatistical interpolation is first leveraged to infer full-resolution topology metrics from the collected sparse measurements. Kriging provides minimum-variance unbiased estimates of spatially correlated variables such as RSRP, interference, attenuation, and SINR by exploiting covariance structure through the empirical semivariogram~\cite{b6,b7}. It is therefore well suited to reconstructing radio environment maps (REMs) from limited samples. Ordinary Kriging (OKG) is used here because of its strong accuracy and manageable computational cost under the propagation conditions of port environments~\cite{Rz,Rm}.

Integrated Access and Backhaul (IAB) offers a second enabling mechanism. Fiber deployment to distributed fixed base stations is costly and disruptive in active logistics areas, while IAB allows access and relay functions to share the same air interface and spectrum~\cite{b10,b11}. Within this framework, fixed IAB donor base stations (FIABs) anchor the topology through wired connectivity, and MIABs provide repositionable access and relay capability that better matches changing traffic and blockage conditions~\cite{b12,b13}.

Existing obstacle-aware deployment methods generally assume that obstacle locations and geometries are known beforehand from maps, site surveys, or ray-tracing databases. In operational smart ports, however, container stacks are continuously rearranged, making such information difficult to maintain and often unavailable at deployment time. This paper addresses this gap by inferring optimization-ready obstacle characterizations directly from the reconstructed REMs. Spatial analysis of the OKG-derived attenuation field reveals high-degradation zones that correspond to physical obstructions. Morphological peak detection~\cite{b18}, spatial grouping, and oriented bounding cuboid fitting are applied to extract compact geometric representations of these zones, adapting established image-analysis techniques~\cite{b14,b18} to an application domain in which, to the best of our knowledge, they have not previously been used.

Building on these ideas, this paper presents \textsc{DOCKING} (\textbf{D}ata-driven \textbf{O}bstacle \textbf{C}haracterization, \textbf{K}riging, and \textbf{I}AB-node positioni\textbf{NG}), a unified pipeline for capacity-aware optimization of an IAB-node deployment in sub-6~GHz smart port environments.

 \textsc{DOCKING} comprises five stages: system modeling, sparse measurement emulation, OKG-based radio map reconstruction, obstacle characterization, and MIAB placement optimization. The approach extends the POSEIDON solution in~\cite{b16} by proposing a measurement-driven approach that removes the need for prior obstacle-geometry, while assuming that FIAB locations and network configuration parameters are known.

The main contributions of this paper are as follows:
\begin{itemize}
\item \textbf{REM-driven obstacle characterization from sparse measurements.}
This paper proposes a REM-driven method that transforms sparse radio measurements into optimization-ready obstacle characterizations for MIAB deployment. The proposed method combines OKG reconstruction, attenuation-region extraction, morphological processing, and oriented cuboid fitting to infer primary blockage structures directly from reconstructed REMs, avoiding the need for prior explicit obstacle-geometry information.
\item \textbf{End-to-end evaluation and experimental demonstration.}
We evaluate the complete \textsc{DOCKING} workflow under realistic smart-port conditions, quantifying REM reconstruction accuracy, obstacle-characterization fidelity, and MIAB capacity gains across multiple deployment scenarios. Additionally, a proof-of-concept (PoC) field campaign demonstrates the practical applicability of the proposed REM-to-obstacle-to-MIAB pipeline by replacing the emulated sparse-measurement stage with real radio measurements.
\end{itemize}

To the best of our knowledge, this is the first framework to derive optimization-ready obstacle abstractions directly from reconstructed REMs, removing dependence on prior obstacle databases and explicit environment geometry.

The remainder of this paper is organized as follows. Section~II reviews related work. Section~III presents the system model and ground-truth radio metric construction. Section~IV describes the OKG reconstruction and obstacle characterization methods. Section~V details the MIAB placement optimization stage. Section~VI presents the performance evaluation. Section~VII concludes the paper.

\section{RELATED WORK}
The problem addressed in this paper intersects four research areas: smart port and industrial wireless networks, REM reconstruction through OKG, obstacle characterization from radio maps, and MIAB deployment optimization.

Smart-port digitalization has received increasing attention as a demanding application for next-generation wireless systems. Potter et al.~\cite{b2} studied barriers and enablers to 5G adoption in ports, including AGV control, predictive maintenance, and cargo tracking, but did not address dynamic blockage-aware planning. Rost et al.~\cite{b3} demonstrated network slicing in a live seaport deployment, but assumed a fixed wired topology and did not consider adaptive relay deployment. Mahmood et al.~\cite{R3} surveyed industrial wireless technologies, but did not study the interaction between dynamic obstacles and deployment optimization in smart seaports.

REM reconstruction from sparse measurements is well established, and OKG remains one of the most widely used approaches. Xia et al.~\cite{b6} proposed adaptive OKG with affinity propagation clustering, but did not address multi-metric reconstruction or deployment optimization. Gao and Fujii~\cite{b7} studied Kriging under adversarial conditions using trusted reference nodes, but did not consider dynamic obstacles or optimization integration. Maeng et al.~\cite{b8} extended Kriging to 3-D environments using UAV data, confirming its usefulness in complex spatial settings. Alternative methods such as DeepREM~\cite{R7} and matrix completion~\cite{R8} highlight the trade-off between interpolation and learning-based reconstruction. In contrast, \textsc{DOCKING} uses OKG to jointly reconstruct RSRP, interference, SINR, and attenuation maps and then feeds them into obstacle characterization and MIAB optimization.

Obstacle characterization from radio propagation data remains emerging. Prior work has mainly focused on coverage-hole detection, clustering, or surface extraction rather than explicit obstacle geometry. Li et al.~\cite{R14} used DBSCAN for signal-region clustering, and Chen et al.~\cite{R16} applied DBSCAN to 3-D point-cloud boundary detection. Related image-processing methods have also been used by Lalak et. al. for object detection and structural characterization~\cite{R15}. Unlike these approaches, \textsc{DOCKING} infers obstacle geometry directly from kriged attenuation maps using morphological peak detection, spatial grouping, and oriented bounding cuboid fitting.

IAB has motivated extensive work on backhaul architecture, scheduling, and topology optimization. Madapatha et al.~\cite{b10} analyzed multi-hop IAB in dense urban and suburban scenarios, but not in obstacle-rich industrial environments and in FR1. Additionally, Madapatha et al.~\cite{R10} proposed a genetic-algorithm framework for joint IAB placement and routing, but used stochastic blockage models rather than measurement-driven reconstruction. Pagin et al.~\cite{b13} and Monteiro et al.~\cite{b12} highlighted the importance of dynamic deployment and resource management in MIAB systems. Geographically aware planning was further studied by Al-Dhuhouri et al.~\cite{R13}. In contrast, \textsc{DOCKING} integrates OKG-derived radio maps, inferred obstacle geometry, and capacity-aware MIAB placement in a unified pipeline for smart port environments.

\section{SYSTEM MODEL AND GROUND-TRUTH RADIO METRIC CONSTRUCTION}
\label{sec_3}
This section defines the seaport environment, network entities, and obstacle
geometry, and constructs the model-based ground-truth radio metric dataset --- a
noise-free record of received signal strength, interference, SINR, and path
attenuation over a discretized two-dimensional grid --- that serves as the
reference for all subsequent pipeline stages. These correspond to Stages~1
and~2 of the \textsc{DOCKING} pipeline (Fig.~\ref{fig:flow}).

\subsection{Environment Geometry and Network Entities}
The study area is discretized into a uniform grid of square pixels with side length $\Delta = 25$~m. Let $\mathcal{G} = \{(x_j, y_j, z_j)\}_{j=1}^{N}$ denote the set of grid points, where $z_j = h_{\mathrm{UE}} = 1.5$~m represents the user equipment (UE) height. Each grid point corresponds to the Euclidean coordinates of the center of a pixel.
A total of $F$ FIABs are deployed at known coordinates
$\mathbf{k}_k = (x_k,y_k,h_{\mathrm{F}})$ with $h_{\mathrm{F}}=10$~m. No MIAB is present at this stage. The carrier frequency for FIAB access links is
$f_{\mathrm{F}}=3.6192$~GHz, and all transmitters radiate with power $P_{\mathrm{t}}=4$~W.

\subsection{Obstacle Representation}
The seaport contains $O$ metallic containers, each modeled as a cuboid defined by
eight vertices $\mathbf{v}_{i,n}\in\mathbb{R}^3$, $n=1,\dots,8$. For each obstacle $i$, the
axis-aligned bounds are computed as (\ref{eq:bounds_x}).
\begin{equation}
x_i^{\min}=\min_n v_{i,n}^{(x)},\quad
x_i^{\max}=\max_n v_{i,n}^{(x)},
\label{eq:bounds_x}
\end{equation}

Analogous expressions can be obtained for $y$ and $z$ coordinates. These bounds allow efficient determination of
line-of-sight (LoS) obstruction between any transmitter–receiver pair.

\subsection{LoS/NLoS Determination}
For each grid point $j$ and FIAB $k$, the segment connecting $\mathbf{k}_k$ and
$\mathbf{j}_j$ is tested for intersection with all obstacle faces. Let
$\mathbf{d}_{jk}=\mathbf{j}_j-\mathbf{k}_k$ denote the direction vector. For each obstacle face, a parametric intersection test of the form in (\ref{eq:parametric_line}) is performed.
\begin{equation}
\mathbf{p}(t)=\mathbf{k}_k + t\,\mathbf{d}_{jk},\qquad t\in[0,1],
\label{eq:parametric_line}
\end{equation}

If any intersection lies within the face boundaries, the link is classified as
non-line-of-sight (NLoS); otherwise, it is LoS. This yields a binary LoS indicator
$\chi_{jk}\in\{0,1\}$ for every grid location and FIAB link.

\subsection{Path Loss Modeling}

Propagation follows the 3GPP UMi Street-Canyon model. For a link with
3D distance $d_{jk}$ and 2D distance $d_{jk}^{\mathrm{2D}}$, the
breakpoint distance is given by \eqref{eq:dbp}, where $c$ is the speed of light in vacuum.
\begin{equation}
d_{\mathrm{BP}} =
\frac{4(h_{\mathrm{F}}-1)(h_{\mathrm{UE}}-1)f_{\mathrm{F}}}{c}
\label{eq:dbp}
\end{equation}

The LoS and NLoS path-loss expressions are given by
\eqref{eq:LoS} and \eqref{eq:pl_nlos}, respectively.
\begin{equation}
\mathrm{PL}_{\mathrm{LoS}}(d_{jk}) =
\begin{cases}
10^{3.24} d_{jk}^{2.1} f_{\mathrm{F}}^{2},
& d_{jk}^{\mathrm{2D}} \le d_{\mathrm{BP}}
\\
10^{3.24}
\frac{d_{jk}^{4} f_{\mathrm{F}}^{2}}
{\bigl(d_{\mathrm{BP}}^{2}
+ (h_{\mathrm{F}}-h_{\mathrm{UE}})^2\bigr)^{0.95}},
& d_{jk}^{\mathrm{2D}} > d_{\mathrm{BP}}
\end{cases}
\label{eq:LoS}
\end{equation}

\begin{equation}
\mathrm{PL}_{\mathrm{NLoS}}(d_{jk}) =
\frac{10^{2.24}\, d_{jk}^{3.53} f_{\mathrm{F}}^{2.13}}
{10^{0.03(h_{\mathrm{UE}}-1.5)}}
\label{eq:pl_nlos}
\end{equation}

The effective path loss is then obtained according to
\eqref{eq:pl_total}, where $\chi_{jk}$ takes value $1$ under LoS conditions and $0$ otherwise.
\begin{equation}
\mathrm{PL}_{jk} =
\chi_{jk}\,\mathrm{PL}_{\mathrm{LoS}}(d_{jk})
+ (1-\chi_{jk})\,\mathrm{PL}_{\mathrm{NLoS}}(d_{jk})
\label{eq:pl_total}
\end{equation}

\subsection{Received Power, Interference, and SINR}

The received power at grid point $j$ from FIAB $k$ is given by (\ref{eq:rx_power}).
\begin{equation}
R_{jk} = \frac{P_{\mathrm{t}}}{\mathrm{PL}_{jk}}
\label{eq:rx_power}
\end{equation}

The serving FIAB is defined as in (\ref{eq:serving_fiab}) as the one achieving the maximum received power:
\begin{equation}
k^\star = \arg\max_k R_{jk}, \qquad
R_j^{\max} = R_{j k^\star}
\label{eq:serving_fiab}
\end{equation}

Interference is computed in~\eqref{eq:interference} as the sum of all
non-serving FIAB contributions. Thermal noise power is modeled
in~\eqref{eq:noise}, where $N_{\mathrm{dBm}}$ is given
by~\eqref{eq:N_dBm} with $\mathrm{NF} = 5\,\mathrm{dB}$ denoting the
assumed receiver noise figure, $\mu = 1$ the numerology index, and $B=106$ the
number of resource blocks. The resulting SINR is given in~\eqref{eq:sinr}.
\begin{equation}
I_j = \sum_{k \,:\, k \neq k^\star} R_{jk}
\label{eq:interference}
\end{equation}
\begin{equation}
N_{\mathrm{mW}} = 10^{\frac{N_{\mathrm{dBm}}-30}{10}}
\label{eq:noise}
\end{equation}
\begin{equation}
N_{\mathrm{dBm}} = -174 + \mathrm{NF} + 10 \log_{10}\!\left(
B \cdot 2^\mu \cdot 15 \cdot 10^3 \cdot 12 \right)
\label{eq:N_dBm}
\end{equation}
\begin{equation}
\mathrm{SINR}_j = \frac{R_j^{\max}}{I_j + N_{\mathrm{mW}}}
\label{eq:sinr}
\end{equation}

\subsection{Attenuation Metric}

For each FIAB $k$, the attenuation at UE $j$ is defined in (\ref{eq:attenuation_link}) and the combined attenuation map is obtained in (\ref{eq:attenuation_map}) by taking the minimum across all FIABs.
\begin{equation}
A_{jk} = 10\log_{10}\!\left(\mathrm{PL}_{jk}\right)
\label{eq:attenuation_link}
\end{equation}
\begin{equation}
A_j = \min_{k} A_{jk}
\label{eq:attenuation_map}
\end{equation}

\subsection{Ground-Truth Dataset}
\label{subsec:ground_truth}
The complete set of ground-truth radio metrics is stored for every UE at each grid pixel, characterized by its Euclidean coordinates:
\[
\bigl\{ R_j^{\max},\, I_j,\, \mathrm{SINR}_j,\, A_j \bigr\}.
\]

This dataset serves as the reference for the OKG reconstruction, obstacle characterization, and MIAB optimization stages.

\begin{figure*}[t]
    \centering
    \includegraphics[width=\textwidth]{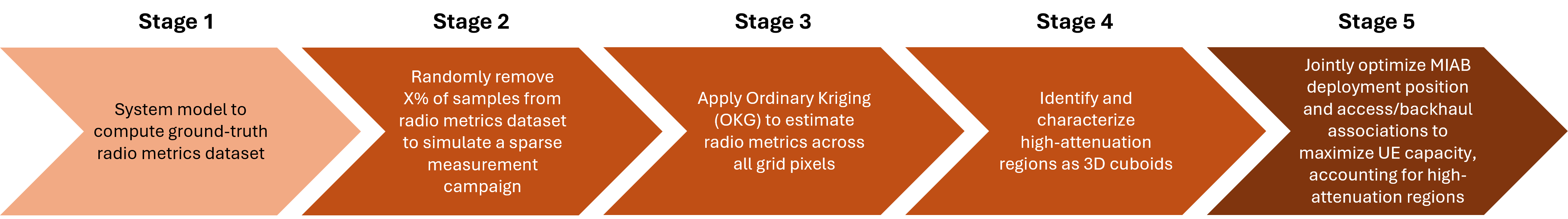}
    \caption{The five-stage \textsc{DOCKING} pipeline: (1)~ground-truth radio
metric construction, (2)~sparse measurement emulation, (3)~OKG-based radio
map reconstruction, (4)~obstacle characterization as 3D cuboids, and
(5)~penalty-augmented GA optimization of MIAB placement and
access/backhaul associations to maximize aggregate UE capacity.}
    \label{fig:flow}
\end{figure*}

\section{SPARSE MEASUREMENT EMULATION, OKG INTERPOLATION, AND HIGH-ATTENUATION REGIONS CHARACTERIZATION}
\label{sec:kriging+clustering}
Building upon the output dataset of Section~\ref{sec_3}, this section
describes the three intermediate stages of the DOCKING pipeline from Fig.~\ref{fig:flow}: (Stage 2) emulation of sparse measurement campaigns through controlled removal of pixel-level radio metrics, (Stage 3)
reconstruction of full-resolution radio maps via OKG, and (Stage 4) characterization of high-attenuation regions as 3D cuboids from the interpolated attenuation field. These stages form the bridge between the theoretical environment model (Stage 1) and the capacity-aware MIAB deployment optimization presented in Section~V (Stage 5).

    \subsection{Sparse Measurement Emulation}
The first intermediate stage, for the theoretical study, emulates a sparse measurement campaign by selectively removing a controlled fraction of the radio metrics computed in Section~\ref{sec_3}. Let $\mathcal{G}$ denote the complete set of grid pixels and let $\rho\in[0,1]$ be the fraction of suppressed measurements. The resulting subset of retained measurement points
$\mathcal{S}\subset\mathcal{G}$ satisfies (\ref{eq:sparsity_relation}).
\begin{equation}
|\mathcal{S}| = (1-\rho)\,|\mathcal{G}| 
\label{eq:sparsity_relation}
\end{equation}
To emulate different spatial sampling behaviors, each grid point $(x_j,y_j,z_j)$ is assigned a
sampling weight derived from a spatial probability density function (PDF) $f(x,y)$, defined
over the normalized coordinates represented in (\ref{eq:normalized_coords}), with $x_{\min},x_{\max},y_{\min},y_{\max}$ denoting the bounds of the study area.
\begin{equation}
\tilde{x} = \frac{x - x_{\min}}{x_{\max} - x_{\min}}, 
\qquad
\tilde{y} = \frac{y - y_{\min}}{y_{\max} - y_{\min}}
\label{eq:normalized_coords}
\end{equation}
Three types of PDFs are considered:
\begin{itemize}
  \item \emph{Uniform sampling} (spatially homogeneous probing) represented by (\ref{eq:uniform_pdf}).
  \begin{equation}
  f_{\mathrm{unif}}(x,y) = 1
  \label{eq:uniform_pdf}
  \end{equation}
  \item \emph{Gaussian sampling} (central concentration of measurements) represented by (\ref{eq:gaussian_pdf}), where $\mu_c = 0.5$ is the normalized center of the domain and $\sigma = 0.25$ controls the spread of the measurement cloud.
\begin{equation}
f_{\mathrm{gauss}}(x,y) =
\exp\!\left(
-\frac{(\tilde{x}-\mu_c)^2 + (\tilde{y}-\mu_c)^2}{2\sigma^2}
\right)
\label{eq:gaussian_pdf}
\end{equation}
  \item \emph{Exponential sampling} (bias towards one side of the area) represented by (\ref{eq:exponential_pdf}), where $\lambda>0$ controls how strongly the sampling is concentrated towards normalized $\tilde{x}=1$ (e.g., the right-hand side of the area, or left-hand if $\tilde{x}=0$).
  \begin{equation}
  f_{\mathrm{exp}}(x,y) =
  \exp\!\bigl(\lambda(\tilde{x}-1)\bigr)
  \label{eq:exponential_pdf}
  \end{equation}
 \end{itemize}
Sampling is performed without replacement, where the probability of selecting a point
$(x_j,y_j)$ is proportional to the chosen PDF $f(x_j,y_j)$, with
$f\in\{f_{\mathrm{unif}},f_{\mathrm{gauss}},f_{\mathrm{exp}}\}$. For each retained pixel
$j\in\mathcal{S}$, the radio metrics included in (\ref{eq:visible_metrics_sparse}) remain in the dataset.
\begin{equation}
\bigl\{ R_j^{\max},\; I_j,\; \mathrm{SINR}_j,\; A_j \bigr\}, \qquad j\in\mathcal{S}
\label{eq:visible_metrics_sparse}
\end{equation}
All other grid points $j\notin\mathcal{S}$ are assigned missing values, producing a sparse
dataset that mimics practical drive-test or walk-test campaigns. This sparse representation constitutes the input to the OKG reconstruction
stage described in the next Section.

\subsection{OKG-Based Radio Metric Reconstruction}
Stage 3 reconstructs full-resolution radio maps from the sparse sampled
measurements resulting from the previous section. For each radio metric 
$m\in\{\mathrm{RSRP},\mathrm{Interference},\mathrm{SINR},\mathrm{Attenuation}\}$, let
$\mathcal{S}_m=\{\mathbf{s}_i=(x_i,y_i)\}$ denote the set of retained samples and let
$p_i^{(m)}$ be the corresponding measured value. The empirical semivariogram is computed
over $N_{\mathrm{bin}}$ distance bins as detailed in (\ref{eq:empirical_variogram}), where $\mathcal{N}(h)$ is the set of sample pairs whose separation distance lies within
the bin centered at lag $h$.
\begin{equation}
\gamma_{\mathrm{emp}}^{(m)}(h)
= \frac{1}{2\,|\mathcal{N}(h)|}
\sum_{(i,j)\in\mathcal{N}(h)}
\bigl(p_i^{(m)} - p_j^{(m)}\bigr)^2
\label{eq:empirical_variogram}
\end{equation}
A theoretical spherical model semivariogram is fitted to $\gamma_{\mathrm{emp}}^{(m)}(h)$ using
nonlinear least squares, according to (\ref{eq:spherical_variogram}). The model parameters are the nugget $n^{(m)}$, sill $c^{(m)}$,
and range $a^{(m)}$.
\begin{equation}
\gamma_{\mathrm{th}}^{(m)}(h)
=
\begin{cases}
n^{(m)} + c^{(m)}
\left(
\dfrac{3h}{2a^{(m)}}+ \right. \\[4pt]
\left. - \dfrac{1}{2}\left(\dfrac{h}{a^{(m)}}\right)^{3}
\right), 
& 0 \le h \le a^{(m)} \\[6pt]
n^{(m)} + c^{(m)}, 
& h > a^{(m)}
\end{cases}
\label{eq:spherical_variogram}
\end{equation}
The spherical model is used because it provides a bounded and physically interpretable correlation range and captures abrupt spatial changes typical of container-port propagation. An isotropic model is assumed, so correlation depends only on separation distance $h$.
For a prediction pixel $\mathbf{s}=(x,y)$, OKG determines the weights $\boldsymbol{\lambda}^{(m)}=[\lambda_1^{(m)},\dots,\lambda_{|\mathcal{S}_m|}^{(m)}]^\top$ and the Lagrange multiplier $\nu^{(m)}$ by solving \eqref{eq:kriging_system}, where the entries of $\boldsymbol{\Gamma}^{(m)}$ and $\boldsymbol{\gamma}^{(m)}$ are given by matrix equation \eqref{eq:gamma_matrix} and vector equation \eqref{eq:gamma_vector}.
\begin{equation}
\begin{bmatrix}
\boldsymbol{\Gamma}^{(m)} & \mathbf{1} \\
\mathbf{1}^{\top} & 0
\end{bmatrix}
\begin{bmatrix}
\boldsymbol{\lambda}^{(m)} \\
\nu^{(m)}
\end{bmatrix}
=
\begin{bmatrix}
\boldsymbol{\gamma}^{(m)}(\mathbf{s}) \\
1
\end{bmatrix}
=
\begin{bmatrix}
\boldsymbol{\gamma}^{(m)}(\mathbf{s}) \\
1
\end{bmatrix}
\label{eq:kriging_system}
\end{equation}
\begin{equation}
\Gamma^{(m)}_{ij}
= \gamma_{\mathrm{th}}^{(m)}\!\left(\|\mathbf{s}_i-\mathbf{s}_j\|\right)
\label{eq:gamma_matrix}
\end{equation}
\begin{equation}
\gamma^{(m)}_i(\mathbf{s})
= \gamma_{\mathrm{th}}^{(m)}\!\left(\|\mathbf{s}-\mathbf{s}_i\|\right)
\label{eq:gamma_vector}
\end{equation}
The OKG predictor of metric $m$ at $\mathbf{s}$ is indicated by \eqref{eq:kriging_prediction} and the associated Kriging variance is given by \eqref{eq:kriging_variance}.
\begin{equation}
\hat{p}^{(m)}(\mathbf{s})
= \sum_{i=1}^{|\mathcal{S}_m|} \lambda_i^{(m)}\,p_i^{(m)}
\label{eq:kriging_prediction}
\end{equation}
\begin{equation}
\sigma_m^2(\mathbf{s})
= \sum_{i=1}^{|\mathcal{S}_m|} \lambda_i^{(m)}\,\gamma^{(m)}_i(\mathbf{s})
+ \nu^{(m)}
\label{eq:kriging_variance}
\end{equation}
To improve numerical stability, the system in (\ref{eq:kriging_system}) is solved with the Moore--Penrose pseudoinverse~\cite[p.~290]{R17}.
The prediction grid is processed in batches of fixed size (e.g., $1000$ points per iteration) to control
memory usage. This procedure yields full-resolution reconstructed maps for all enabled
metrics in $\{\mathrm{RSRP},\mathrm{Interference},\mathrm{SINR},\mathrm{Attenuation}\}$ dataset.

\subsection{High-Attenuation Regions Characterization}
\label{subsec:Clustering}
Stage 4 identifies high-attenuation regions and approximates them as 3D cuboids. The choice of cuboidal obstacle characterization is motivated by both environmental and methodological considerations. First, container terminals are predominantly composed of shipping-container stacks whose physical structure is inherently cuboidal, making cuboids a natural abstraction of the dominant blockage sources. Second, this choice leverages previous work by the authors, where geometric obstruction effects can be accurately captured through ray-intersection analysis using cuboidal obstacles. Consequently, the inferred cuboids preserve compatibility with established line-of-sight and blockage calculations while providing a compact and optimization-friendly characterization of attenuation-inducing structures.

\emph{1) Attenuation field formation:}
OKG produces a dense attenuation field $A(x,y)$ with no missing entries.

\emph{2) Normalization:}
The attenuation field is normalized to the interval $[0,1]$ using \eqref{eq:normalization}, where $A_{\min}$ and $A_{\max}$ denote the minimum and maximum attenuation values over the grid.
\begin{equation}
A_{\mathrm{norm}}(x,y)
= \frac{A(x,y)-A_{\min}}{A_{\max}-A_{\min}}
\label{eq:normalization}
\end{equation}

\emph{3) Peak extraction:}
High-attenuation regions are detected with an $H$-maxima transform, as defined in \eqref{eq:hmaxima}, which retains only those pixels whose attenuation values exceed all neighboring pixels within their local neighborhood by at least a threshold $h_{\mathrm{val}}\%$ (typical 0.15). This operation suppresses broad plateaus and low-contrast variations, isolating only \emph{salient peaks}, consistent with the presence of dominant obstructions such as container stacks.
\begin{equation}
M(x,y)
= \mathrm{Hmax}\!\left(A_{\mathrm{norm}}(x,y),\,h_{\mathrm{val}}\right)
\label{eq:hmaxima}
\end{equation}

\emph{4) Morphological expansion:}
To account for the spatial extent of the detected attenuation peaks, a morphological dilation is applied, as per \eqref{eq:dilation}, where $D$ is a disk-shaped structuring element. This choice ensures isotropic expansion of the detected peaks, avoiding directional bias while producing compact and spatially coherent regions. As a result, isolated high-attenuation pixels are transformed into contiguous areas that better reflect the physical extent of underlying obstructions (e.g., container stacks). This step is essential to stabilize subsequent geometric extraction, as it provides well-defined regions without imposing prior assumptions on obstacle orientation that follows.
\begin{equation}
M_{\mathrm{dil}}(x,y)
= M(x,y) \oplus D
\label{eq:dilation}
\end{equation}

\emph{5) Region extraction:}
Connected-component analysis is then performed on $M_{\mathrm{dil}}(x,y)$ to identify $O$ disjoint regions
$\{\mathcal{C}_1,\dots,\mathcal{C}_O\}$. Each region corresponds to a contiguous set of high-attenuation pixels. For each $\mathcal{C}_i$, second-moment analysis is used to extract its geometric descriptors, namely the centroid $\mathbf{c}_i$, the principal axis lengths $L_i$ and widths $W_i$, and the orientation $\theta_i$, which collectively define an equivalent elliptical footprint of the region.

\emph{6) Geometric reconstruction:}
Each detected region is approximated by a rotated rectangular base aligned with its principal axes. The corner coordinates are obtained as in \eqref{eq:rotated_corners}, where the rotation matrix is defined by \eqref{eq:rotation_matrix}, and the offset vectors satisfy \eqref{eq:rectangle_offsets}.
\begin{equation}
\mathbf{q}_{i,\ell}
= \mathbf{c}_i + \mathbf{R}(\theta_i)\,\mathbf{u}_\ell,
\quad \ell=1,\dots,4
\label{eq:rotated_corners}
\end{equation}
\begin{equation}
\mathbf{R}(\theta_i) =
\begin{bmatrix}
\cos\theta_i & -\sin\theta_i \\
\sin\theta_i & \cos\theta_i
\end{bmatrix}
\label{eq:rotation_matrix}
\end{equation}
\begin{equation}
\mathbf{u}_\ell \in
\left\{
\left[-\tfrac{L_i}{2},-\tfrac{W_i}{2}\right],
\left[\tfrac{L_i}{2},-\tfrac{W_i}{2}\right],
\left[\tfrac{L_i}{2},\tfrac{W_i}{2}\right],
\left[-\tfrac{L_i}{2},\tfrac{W_i}{2}\right]
\right\}
\label{eq:rectangle_offsets}
\end{equation}

\emph{7) Height assignment and 3D extrusion for cuboid construction:}
The vertical extent of each obstacle is modeled uniformly as \eqref{eq:height_model}, where $h_{\min} = 2.59\,\mathrm{m}$ denotes the height of a standard
ISO shipping container and $N_i$ is the number of stacked units, drawn
uniformly from $N_{\min} = 2$ to $N_{\max} = 5$ to reflect the typical
stacking range observed in container terminals. Because height information
is not recoverable from 2D radio measurements alone, this stochastic
assignment provides a statistically representative ensemble of obstacle
heights without requiring vertical sensing. The sensitivity of downstream
LoS computations to this choice is limited in practice, since the UE height
($1.5\,\mathrm{m}$) is well below the minimum stack height
($N_{\min} \cdot h_{\min} = 5.18\,\mathrm{m}$), ensuring NLoS conditions
are preserved regardless of the sampled $N_i$. Given the base rectangle $\{\mathbf{q}_{i,\ell}\}_{\ell=1}^{4}$, the cuboid is obtained by vertical extrusion, with the bottom face at $z=0$ and the top face at $z=H_i$, yielding an eight-vertex cuboid representation, as per \eqref{eq:cuboid_vertices}.
\begin{equation}
H_i = N_i\,h_{\min}, \qquad
N_i \sim \mathcal{U}\!\left(N_{\min},N_{\max}\right)
\label{eq:height_model}
\end{equation}
\begin{equation}
\mathbf{v}_{i,\ell}
=
\begin{cases}
\bigl(\mathbf{q}_{i,\ell},\,0\bigr), & \ell=1,\dots,4 \\[4pt]
\bigl(\mathbf{q}_{i,\ell-4},\,H_i\bigr), & \ell=5,\dots,8
\end{cases}
\label{eq:cuboid_vertices}
\end{equation}
Each region is finally represented as a 3D cuboid defined by eight vertices, forming the inferred obstacle set, as in \eqref{eq:obstacle_set}.
\begin{equation}
\widehat{\mathcal{O}}
= \left\{
\{\mathbf{v}_{i,\ell}\}_{\ell=1}^{8}
\right\}_{i=1}^{O}
\label{eq:obstacle_set}
\end{equation}
These cuboids provide a compact three-dimensional approximation of the dominant high-attenuation regions induced by physical obstructions. Importantly, they are inferred solely from sparse sampled measurements and OKG reconstruction, without requiring explicit knowledge of the environment layout, and this construction enables direct integration of obstacle geometry into subsequent propagation analysis of MIAB deployment optimization.

\section{BACKHAUL-AWARE MIAB PLACEMENT AND ASSOCIATION OPTIMIZATION}
\label{sec:poseidon-plus}
The proposed optimization stage extends the capacity-aware placement
framework of~\cite{b16} to a macro-time-evolving setting in which UE
positions, MIAB placement, its backhaul, and associations are re-optimized at
each of $T$ trajectory snapshots.

\subsection{Pixel-Grid Discretization and UE Mobility}
The study area is discretized into a uniform pixel grid of resolution $\Delta_g$ (e.g., $25\,\text{m}\times25\,\text{m}$). At each time step $t\in\{1,\dots,T\}$, UE $j$ follows a 2-D random walk described in~\eqref{eq:random_walk}, where $v$ is the step size and $\mathbf{d}_j^{(t)}$ is a random unit direction resampled whenever the candidate position falls inside an obstacle footprint. The accepted position is snapped to the closest valid grid point, as defined in~\eqref{eq:snap}.
\begin{equation}
\mathbf{p}_j^{(t)} = \mathbf{p}_j^{(t-1)} + v\,\mathbf{d}_j^{(t)},
\label{eq:random_walk}
\end{equation} 
\begin{equation}
\hat{\mathbf{p}}_j^{(t)} =
\underset{\mathbf{q}\in\mathcal{G}\setminus\mathcal{P_O}}{\arg\min}
\|\mathbf{p}_j^{(t)}-\mathbf{q}\|_2.
\label{eq:snap}
\end{equation}  

\subsection{Kriging-Based SINR Maps and Inter-FIAB Interference}
\label{subsection:kriging_model}
Channel quality for FIAB-related links (UE--FIAB access and
MIAB--FIAB backhaul) is derived from the sparse sampled emulated measurement campaigns in Stage~2 and reconstructed via OKG in Stage~3. For each pixel $\mathbf{q}$ and FIAB $k$,
the field-measured RSRPs $\{P_{k}(\mathbf{q}_s)\}$ at sampled
locations $\mathbf{q}_s$ are interpolated via OKG to yield a dense
SINR map. The per-pixel SINR explicitly captures inter-FIAB
interference as defined in~\eqref{eq:sinr_fiab}, where
$N_0 = -174 + \text{NF} + 10\log_{10}(B\,\Delta_F)$\,[dBm]
is the thermal noise power, NF is the $5\,\text{dB}$ noise
figure, $B$ is the number of resource blocks, and
$\Delta_F = 2^{\mu}\cdot15\,\text{kHz}\cdot12$ is the per-RB
bandwidth for numerology $\mu$.
\begin{equation}
  \gamma_{jk}^{F} =
    \frac{P_{k}\!\left(\hat{\mathbf{p}}_j^{(t)}\right)}
         {\displaystyle\sum_{k'\neq k}
         P_{k'}\!\left(\hat{\mathbf{p}}_j^{(t)}\right) + N_0}
  \label{eq:sinr_fiab}
\end{equation}
Because the OKG-reconstructed measurement maps already embed
obstacle-induced shadowing and inter-cell interference, no
explicit knowledge of the obstacle geometry is required for
FIAB-related links. Analogous OKG-interpolated SINR values,
computed at the MIAB antenna height, are used for the
MIAB--FIAB backhaul link as defined in~\eqref{eq:sinr_bh},
where $\hat{\mathbf{p}}_m^{(t)}$ is the best found MIAB position
at snapshot $t$.
\begin{equation}
  \gamma_{mk}^{BH} =
    \left.\gamma^{F}\right|_{\mathbf{q}\,=\,\hat{\mathbf{p}}_m^{(t)}}
  \label{eq:sinr_bh}
\end{equation}

Because the MIAB position is a decision variable, the analytical 3GPP UMi Street-Canyon model is used for MIAB-to-UE access links. Obstacle blockage is detected through the cuboids identified in Stage 4, using the same intersection test as in~\cite{b16}.

\subsection{Penalty-Augmented Optimization Within a Real/Integer-Coded Genetic Algorithm (GA)}
The spectral efficiency, per-UE throughput, and overhead model
follow the same formulation as in~\cite{b16}, retaining the 3GPP
MCS Table~2 upper-bound clipping ($\Gamma_{\max}=7.4063$\,bps/Hz),
the TDD PHY-MAC overhead ($\epsilon_{\rm PHY}=0.1973$), and the
access- and backhaul-layer protocol overhead fractions
($\epsilon_{\rm acc}=0.0267$, $\epsilon_{\rm bh}=0.0532$). These
overhead values stem from the calibration of the model with
OpenAirInterface performed in~\cite{b16}.
The FIAB downlink capacity is given by~\eqref{eq:cap_fiab_plus},
where $\eta(\cdot)$ is the clipped spectral efficiency defined
in~\cite{b17}, $\Delta_F = 2^{\mu}\!\cdot\!15\,\text{kHz}\!\cdot\!12$
is the per-RB bandwidth with numerology $\mu=1$, and
\eqref{eq:load} represents the total load on FIAB $k$, accounting
for both directly associated UEs and UEs relayed through the MIAB
($U_m$). Weighted Round Robin (WRR) scheduling is adopted because it provides
proportionally fair resource block allocation among heterogeneous UE sets
while remaining computationally tractable for real-time
operation~\cite{ietf-wfq,b16}, with overhead parameters calibrated against
OpenAirInterface measurements as detailed in~\cite{b16}.
\begin{equation}
  C_{jk}^{F} =
    (1-\epsilon_{\rm PHY})(1-\epsilon_{\rm acc})\,    B\,\Delta_F\,\frac{\eta\!\left(\gamma_{jk}^{F}\right)}{U_k^z}
  \label{eq:cap_fiab_plus}
\end{equation}
\begin{equation}
  U_k^z = \sum_{j=1}^{N} s_{jk}^{F} + U_m \cdot s_{mk}^{F}
  \label{eq:load}
\end{equation}
The MIAB access capacity $C_{jm}^{M}$ and backhaul capacity
$C_{mk}^{BH}$ are computed analogously, replacing $\Delta_F$ with
the MIAB numerology bandwidth $\Delta_M$ and $\epsilon_{\rm acc}$
with $\epsilon_{\rm bh}$ where appropriate, as detailed
in~\cite{b16}.

Stage~5 maximizes the aggregate downlink capacity of the special-team
subset — a designated group of $N_s$ UEs drawn from the $N$ UEs present, representing a mission-critical or priority service team — over the joint MIAB placement, UE--cell, and MIAB--FIAB backhaul associations. The principal
departure from~\cite{b16} is the shift to a \emph{solver-based}
GA formulation, in which all constraints are absorbed into a
quadratic penalty term.

The GA minimizes the fitness function $f(\mathbf{x})$ in
\eqref{eq:fitness}, which combines the negative aggregate capacity
of the $N_s$ UEs with the penalty term $\Pi(\mathbf{x})$ in
\eqref{eq:penalty}. The penalty coefficients $\lambda_1=10^{10}$, $\lambda_2=10^{8}$, and
$\lambda_3=10^{3}$ are selected to impose a strict hierarchy among constraints, such that violations of higher-priority constraints dominate those of lower-priority ones in the fitness evaluation.

Their magnitudes are set to exceed the maximum achievable aggregate capacity
(on the order of $10^{7}$--$10^{8}$\,bps for the scenarios considered) by
at least one order of magnitude at each priority level, so that any
constraint violation incurs a fitness cost larger than the entire feasible
objective range. This ensures that the GA always prefers a feasible solution
of lower capacity over an infeasible one of higher capacity.
$\lambda_1$ penalizes UE multi-association (highest priority),
$\lambda_2$ penalizes backhaul violations and capacity overflow
(intermediate priority), and $\lambda_3$ penalizes soft radio constraints
(lowest priority).
\begin{equation}
  f(\mathbf{x}) =
    -\sum_{j=1}^{N_s}\!\left(
      \sum_{k=1}^{F} C_{jk}^{F}\, s_{jk}^{F}
      + \sum_{m=1}^{M} C_{jm}^{M}\, s_{jm}^{M}
    \right)
    + \Pi(\mathbf{x})
  \label{eq:fitness}
\end{equation}
\begin{align}
  \Pi(\mathbf{x}) &=
    \lambda_1 \sum_{j}
      \Bigl(\sum_{k} s_{jk}^{F} + \sum_{m} s_{jm}^{M} - 1\Bigr)^{\!2}
    \notag \\
  &+\, \lambda_2 \sum_{m}
      \Bigl(\sum_{k} s_{mk}^{F} - 1\Bigr)^{\!2}
    \notag \\
  &+\, \lambda_2 \sum_{m}
      \Bigl[\max\!\Bigl(0,\,
        \sum_{j} C_{jm}^{M} s_{jm}^{M}
        - \sum_{k} C_{mk}^{BH} s_{mk}^{F}
      \Bigr)\Bigr]^{\!2}
    \notag \\
  &+\, \lambda_3 \sum_{j,m}
      \bigl[\max\bigl(0,\, qRxLevMin - R_{jm}^{\rm dBm}\bigr)\bigr]\,
      s_{jm}^{M}
    \notag \\
  &+\, \lambda_3 \sum_{j,m}
      \bigl[\max\bigl(0,\, 10 - d_{2D,jm}\bigr)\bigr]\,
      s_{jm}^{M}
    \notag \\
  &+\, \lambda_2\sum_{m}
      \mathbf{1}\!\left[\mathbf{q}_m \in \mathcal{P_O}\right]
  \label{eq:penalty}
\end{align}
The first three lines of~\eqref{eq:penalty} enforce: i) single-cell
association per UE; ii) a single backhaul FIAB per MIAB; and iii)
backhaul capacity sufficiency. The remaining lines penalize
violations of the minimum RSRP constraint
($R_{jm}^{\mathrm{dBm}} \geq q_{\mathrm{RxLevMin}}$)~\cite{b11-a},
the minimum 2D separation constraint ($d_{2D,jm} \geq 10\,\mathrm{m}$),
and the obstacle exclusion constraint
($\mathbf{q}_m \notin \mathcal{P_O}$) for MIAB placement. A
fast-convergence mode reduces the GA stall-generation limit and
the maximum number of generations, at the expense of reduced
optimality. The optimization is independently performed at each
of the $T$ snapshots, yielding the solution $\mathbf{x}^*$
returned by GA, including the MIAB placement
$\mathbf{q}_m^*$, the association matrices $s_{jk}^{F*}$,
$s_{jm}^{M*}$, $s_{mk}^{F*}$, and the individual and aggregate
UE capacities. 

\section{PERFORMANCE EVALUATION AND RESULTS}
\label{sec:results}
This section evaluates the complete DOCKING pipeline through simulation-based analysis and a PoC field demonstration. We assess OKG reconstruction accuracy, obstacle characterization fidelity, MIAB capacity gains, and agreement with measurements obtained in a real deployment environment.

\begin{figure*}[t]
    \centering
    \includegraphics[width=\textwidth]{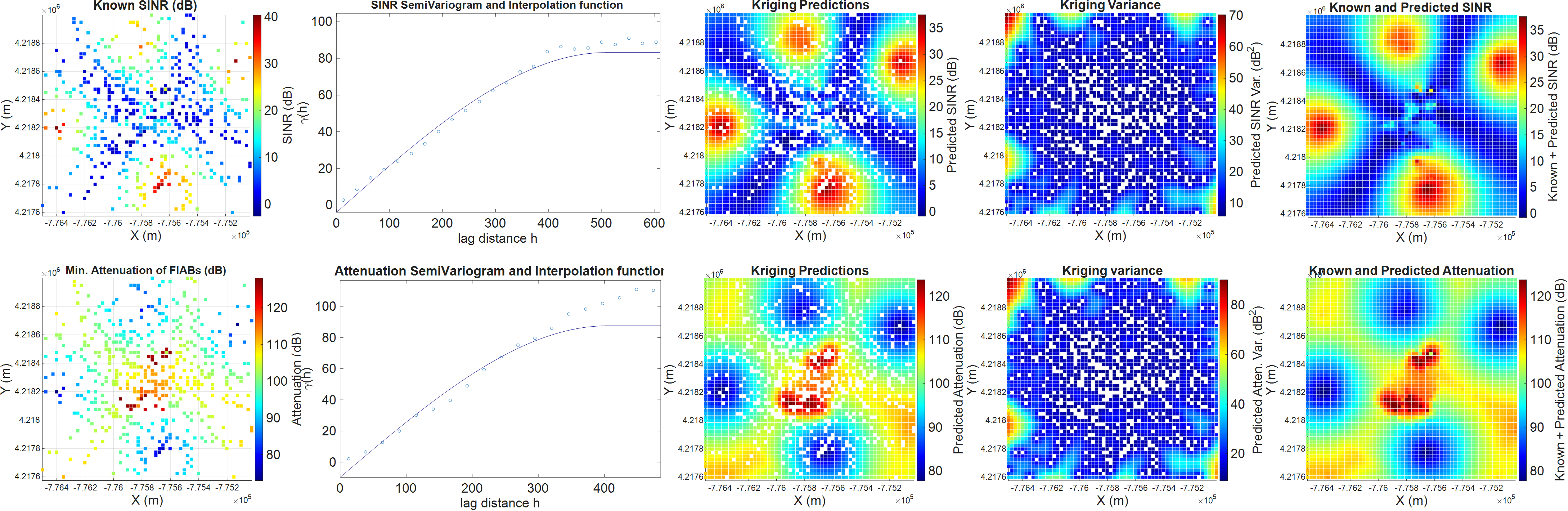}
    \caption{OKG reconstruction results for SINR (top row) and minimum
    attenuation (bottom row). From left to right: emulated measurement campaigns,
    fitted spherical semivariogram, OKG prediction map, OKG
    variance map, and combined known-and-predicted maps.}
    \label{fig:kriging_pic1}
\end{figure*}

\subsection{OKG Interpolation of Radio Metrics}
\label{sec:results_kriging}

The evaluation uses four FIABs deployed over a seaport area discretized into a $25\,\text{m}\times 25\,\text{m}$ pixel grid with ten container-stack obstacles. Measurements cover approximately $15\%$ of pixels, drawn from a two-dimensional Gaussian distribution centered on the domain, producing denser observations toward the interior --- consistent with the practical infeasibility of exhaustive field surveys in an operational seaport.

In Stage~2 of the DOCKING pipeline (Fig.~\ref{fig:flow}), logged RSRP values yield per-pixel SINR and minimum attenuation as defined in~\eqref{eq:sinr} and~\eqref{eq:attenuation_map}. The sparse observations already reveal the dominant spatial structure: SINR spans $0$--$40\,\text{dB}$ with localized high-SINR pockets, while attenuation occupies a narrower $80$--$130\,\text{dB}$ range, reflecting path loss and container-induced shadowing.

OKG with a least-squares-fitted spherical semivariogram (Stage~3) reconstructs a continuous surface over unsampled pixels. The empirical semivariograms confirm well-defined autocorrelation: the SINR variogram reaches its sill at $\approx500\,\text{m}$, while the attenuation variogram saturates earlier at $\approx400\,\text{m}$, consistent with the more localized nature of obstacle-induced shadowing. The resulting prediction maps preserve high-SINR corridors, low-SINR shadow regions, and correctly identify the high-attenuation central port area. OKG variance remains below $70\,\text{dB}^2$ (SINR) and $80\,\text{dB}^2$ (attenuation) across the domain, confirming sufficient interpolation confidence for downstream pipeline stages, as can be observed in Fig.~\ref{fig:kriging_pic1}.

\begin{figure}    \includegraphics[width=\columnwidth]{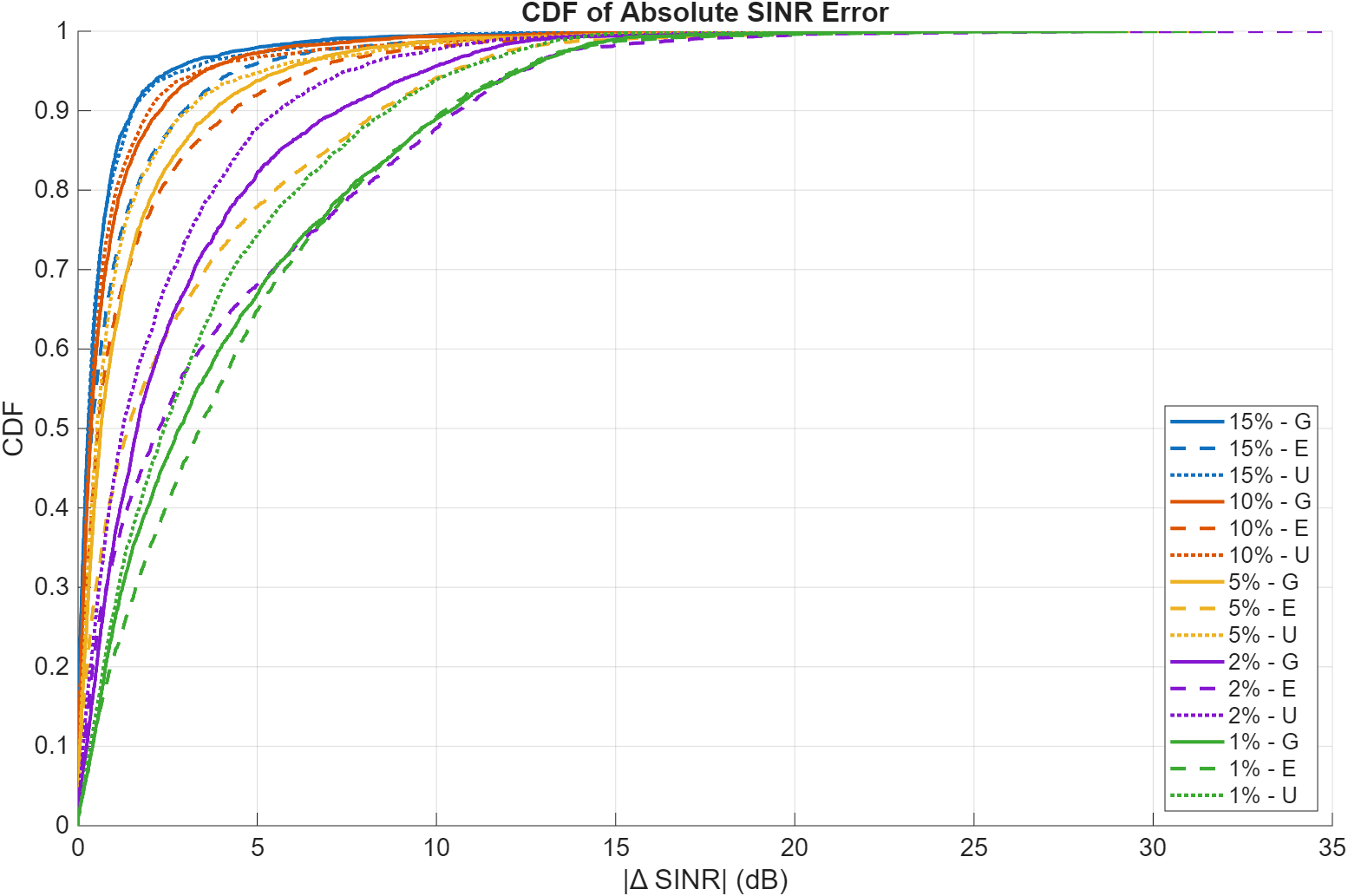}
    \caption{CDFs of the absolute SINR prediction error $|\Delta\text{SINR}|$ for five sampling densities ($1\%$, $2\%$, $5\%$, $10\%$, $15\%$) and three pixel-selection strategies: Gaussian~(G), Exponential~(E), and Uniform~(U).}
    \label{fig:cdf_sinr}
\end{figure}

\begin{figure}    \includegraphics[width=\columnwidth]{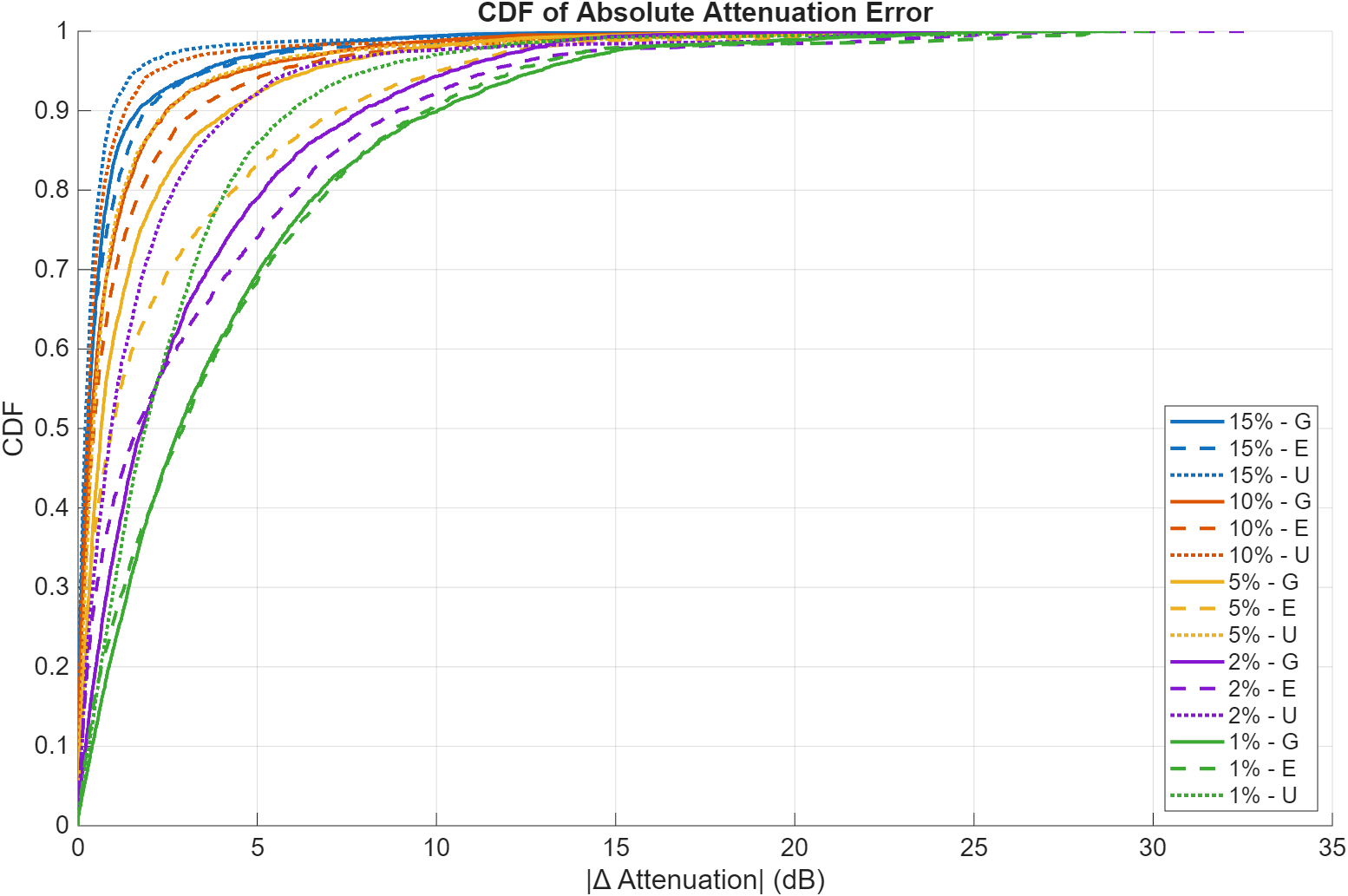}
    \caption{CDFs of the absolute attenuation prediction error $|\Delta\text{Attenuation}|$ for five sampling densities ($1\%$, $2\%$, $5\%$, $10\%$, $15\%$) and three pixel-selection strategies: Gaussian~(G), Exponential~(E), and Uniform~(U).}
    \label{fig:cdf_attenuation}
\end{figure}

To assess robustness, five sampling densities ($15\%$, $10\%$, $5\%$, $2\%$, $1\%$) are combined with three pixel-selection strategies --- Gaussian~(G), Exponential~(E), and Uniform~(U) --- yielding fifteen configurations per metric. Accuracy is quantified via absolute prediction errors $|\Delta\text{SINR}|$ and $|\Delta\text{Attenuation}|$ against the ground-truth dataset of Section~\ref{sec_3}; CDFs are shown in Figs.~\ref{fig:cdf_sinr} and~\ref{fig:cdf_attenuation}.

Three trends emerge consistently. First, sampling density dominates accuracy: at $15\%$, over $90\%$ of pixels show errors below $\approx3\,\text{dB}$ (SINR) and $2\,\text{dB}$ (attenuation); at $1\%$, these rise to $\approx10\,\text{dB}$ and $8\,\text{dB}$. Second, the Gaussian strategy outperforms Exponential and Uniform at all densities due to its higher concentration of samples in the high-variability central area, though differences become negligible at $15\%$. Third, attenuation is interpolated more accurately than SINR in all configurations, owing to its narrower dynamic range and stronger spatial correlation. Overall, Gaussian sampling at $15\%$ --- adopted throughout the remainder of this work --- achieves sub-$3\,\text{dB}$ errors at the $90\%$ percentile for both metrics, sufficient to support the subsequent stages of the DOCKING pipeline.

\subsection{High-Attenuation Region Characterization as 3D Cuboids}
\label{sec:results_clustering}

Stage~4 processes the OKG-reconstructed attenuation map and applies the characterization procedure of Section~\ref{sec:kriging+clustering} to detect high-attenuation regions and approximate them as 3D cuboids. Attenuation is preferred over SINR as it directly reflects obstacle geometry, while SINR also captures interference effects irrelevant to obstacle localization.

\begin{figure}[t]
    \centering
    \includegraphics[width=\columnwidth]{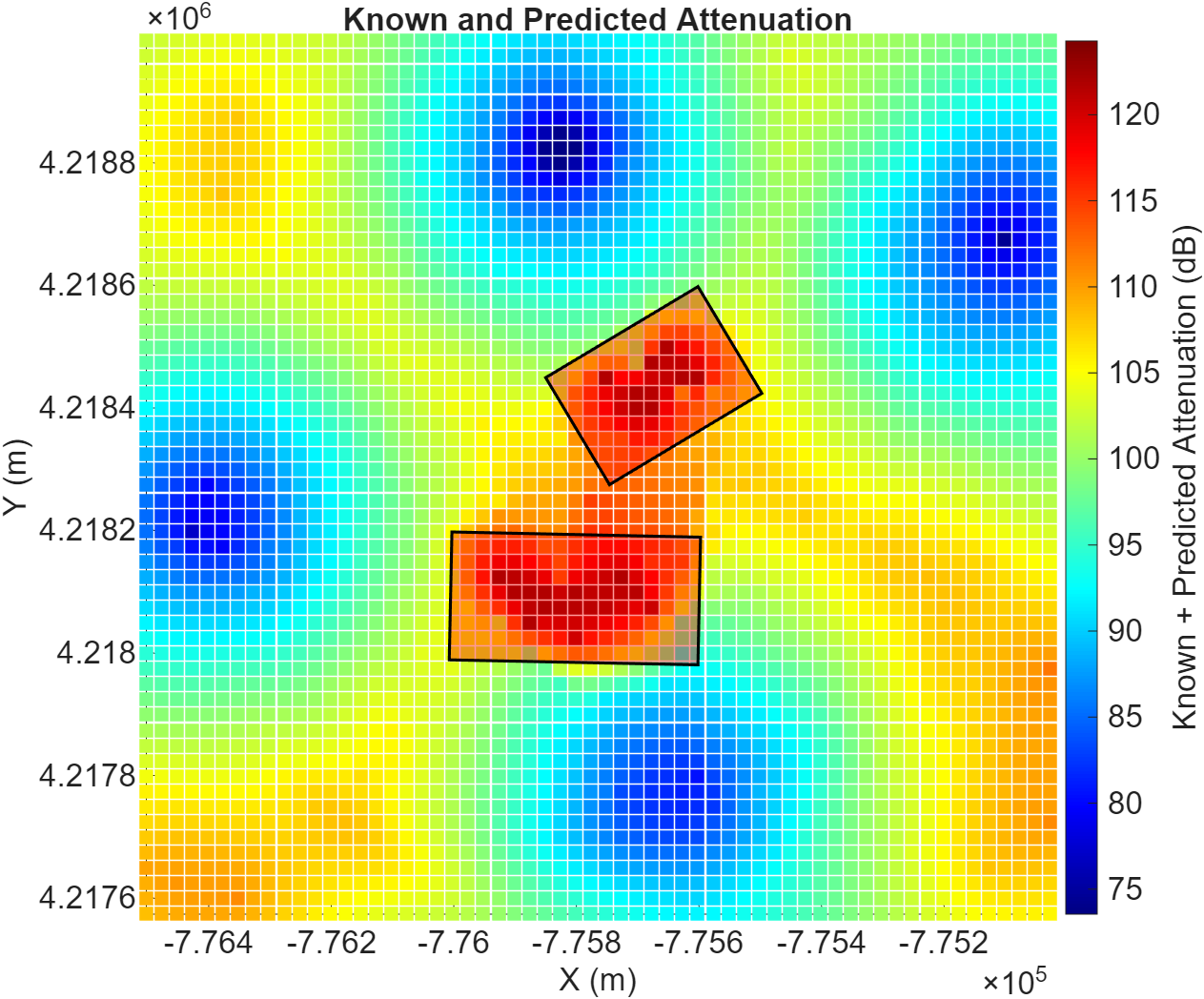}
    \caption{Combined known-and-predicted attenuation map with the fitted
attenuation cuboids (black outlines) overlaid on the dominant
high-attenuation regions in the central port area. The number of inferred
cuboids equals the number of connected components identified in Stage~4.}
    \label{fig:clustering_map}
\end{figure}

Fig.~\ref{fig:clustering_map} illustrates the outcome for the reference scenario (four FIABs, ten obstacles, $25\,\text{m}\times 25\,\text{m}$ grid, $15\%$ Gaussian sampling). The two fitted cuboids (black outlines) closely enclose the dominant high-attenuation regions in the central port area (attenuation $\gtrsim 120\,\text{dB}$), capturing the main obstruction zones without requiring explicit knowledge of container geometry.

\begin{figure}[t]
    \centering
    \includegraphics[width=\columnwidth]{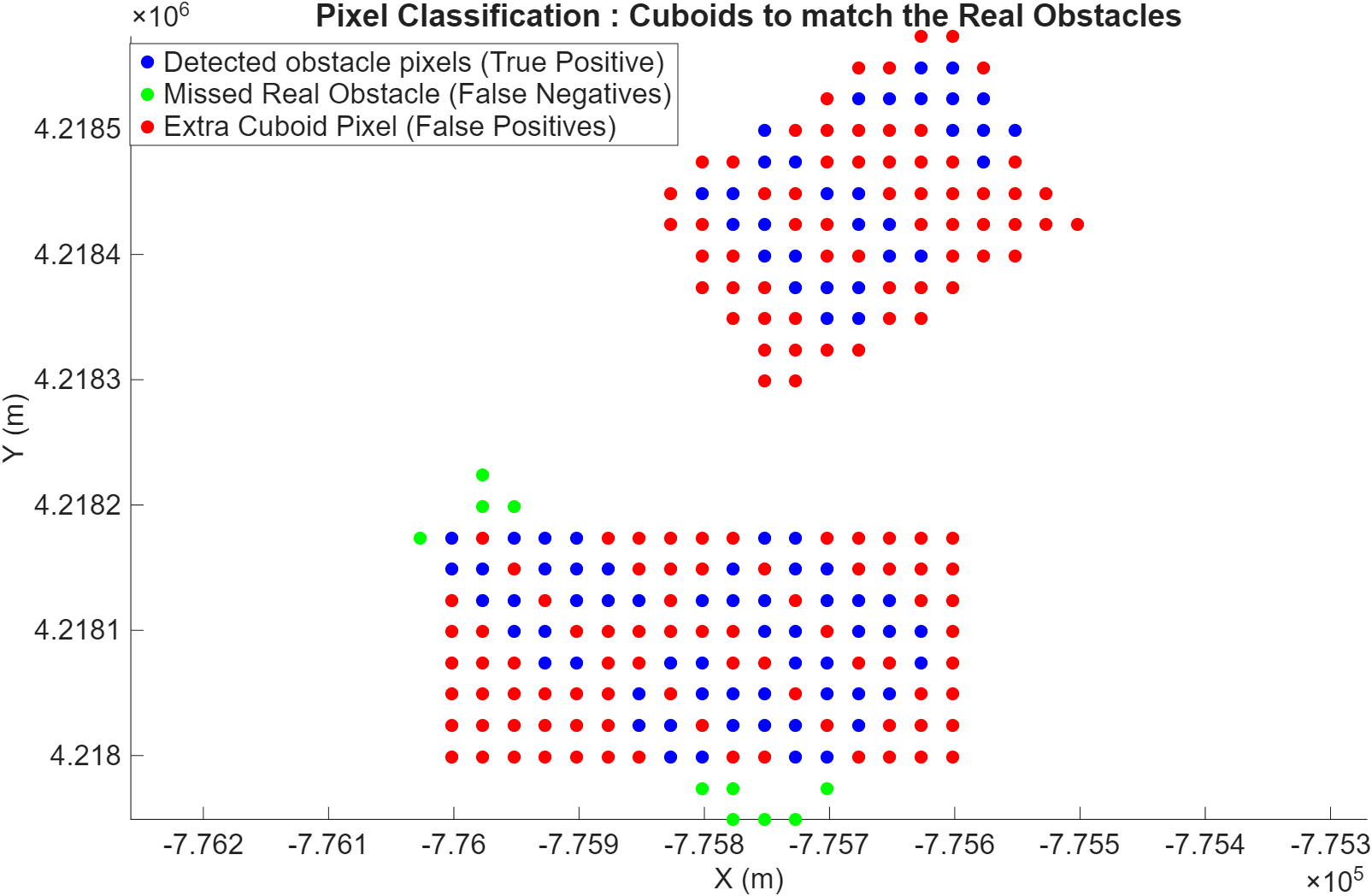}
    \caption{Pixel-level classification of the Stage~4 characterization outcome against the ground-truth obstacle layout: true positive pixels correctly covered by a fitted cuboid (blue), false negative pixels corresponding to missed real obstacle pixels (green), and false positive pixels falling outside the actual obstacle footprint (red).}
    \label{fig:clustering_pixels}
\end{figure}

Fig.~\ref{fig:clustering_pixels} provides a pixel-level comparison against the ground-truth layout. The fitted cuboids achieve approximately $90\%$ true-positive coverage, with false positives concentrated at the corners of the non-axis-aligned upper region and false negatives limited to irregular boundaries and peripheral areas, confirming reliable capture of core obstruction regions.
Parameter selection for Stage~4 involves two interdependent quantities: the attenuation threshold $h_\text{val}$ and the dilation radius applied to the binary mask before connected-component extraction. A practical procedure starts with $h_\text{val} = 0.15$ and single-pixel dilation, accepting the configuration if the true positive rate exceeds $85$--$90\%$ and false positives remain below $5$--$10\%$. The core trade-off is between cuboid count and false positive coverage: a low $h_\text{val}$ produces many small components, increasing solver complexity in Stage~5, whereas a high $h_\text{val}$ or large dilation radius merges regions into fewer but oversized cuboids, inflating false positives at obstacle boundaries as visible in Fig.~\ref{fig:clustering_pixels}. Parameters should therefore be adjusted iteratively until both criteria are jointly satisfied.

\begin{figure}
    \includegraphics[width=\columnwidth]{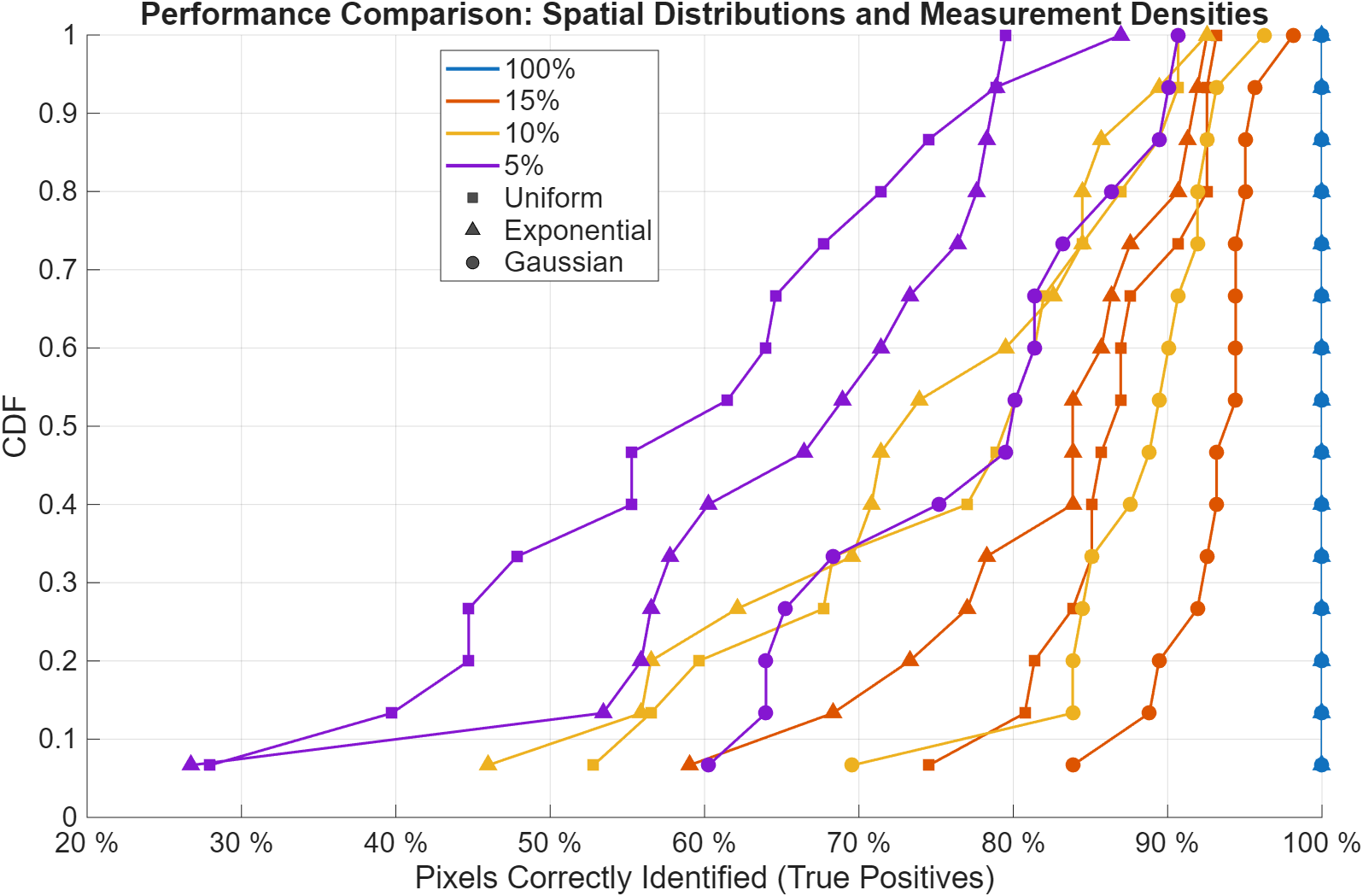}
    \caption{CDFs of the true positive pixel rate for Stage~4 of the DOCKING pipeline, evaluated across four sampling densities ($5\%$, $10\%$, $15\%$, and $100\%$) and three pixel-selection strategies: Uniform~(U), Exponential~(E), and Gaussian~(G).}
    \label{fig:clustering_cdfs}
\end{figure}

Fig.~\ref{fig:clustering_cdfs} presents CDFs of the true positive pixel rate across four sampling densities ($5\%$, $10\%$, $15\%$, $100\%$) and three selection strategies. The $100\%$ case yields a true positive rate of $100\%$ across all realizations, establishing an upper bound. As density decreases, CDFs shift progressively leftward: at $15\%$, median rates range from $88\%$ to $95\%$; at $10\%$, a non-negligible fraction falls below $80\%$, particularly for Exponential sampling; at $5\%$, median rates drop below $70\%$ (Exponential) and $65\%$ (some Uniform realizations), indicating insufficient reconstruction for reliable obstacle localization.

The Gaussian strategy consistently matches or outperforms Uniform sampling due to its concentration of observations in the central port area, while Exponential sampling exhibits the largest spread and heaviest left tail, making it the least reliable choice. Overall, obstacle characterization remains reliable with median true positive rates exceeding $85\%$ at $15\%$ Gaussian sampling, which constitutes a practical lower bound below which attenuation cuboid accuracy may no longer suffice for reliable path-loss and line-of-sight computations in the optimization stage.

\subsection{MIAB-Augmented Capacity Gains}
\label{sec:results_poseidon}
\begin{figure}[!t]
    \centering
    \begin{subfigure}[t]{\columnwidth}
        \centering        \includegraphics[width=\columnwidth]{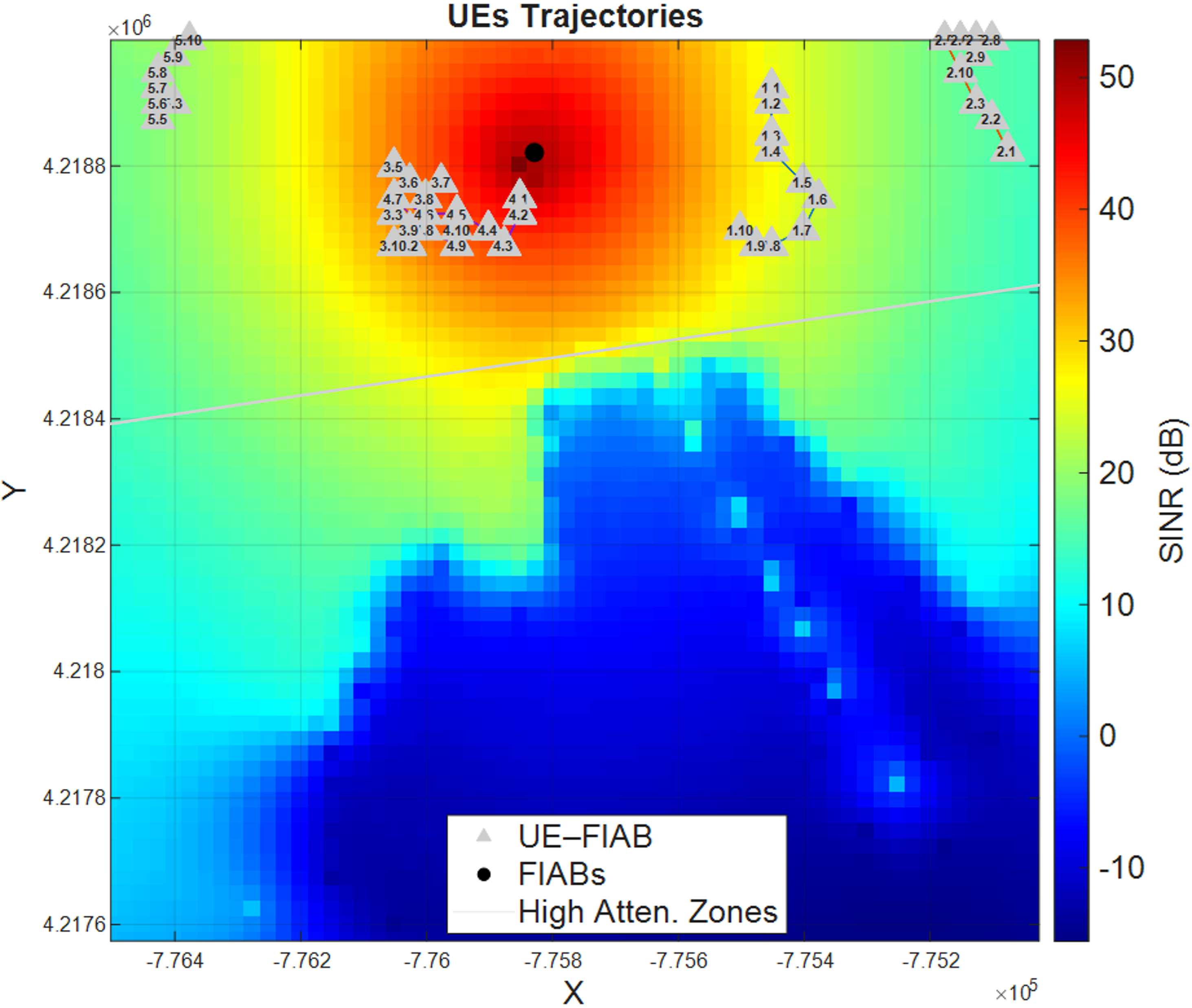}
        \caption{FIAB-only baseline: all UEs associated with the single FIAB (triangle markers) across all $T=10$ snapshots.}
        \label{fig:usecase_fiab}
    \end{subfigure}
    \vspace{2pt}
    \begin{subfigure}[t]{\columnwidth}
        \centering        \includegraphics[width=\columnwidth]{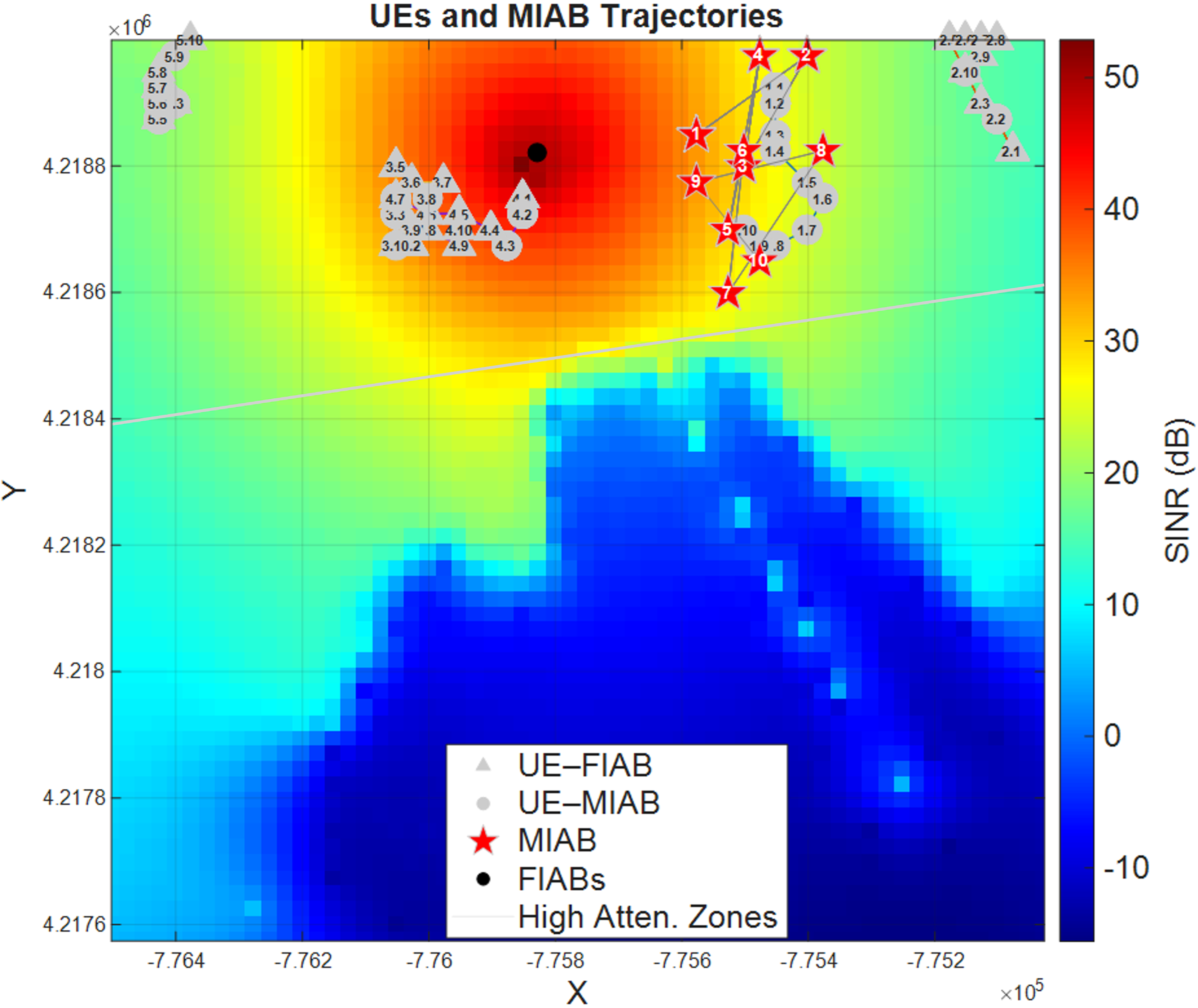}
        \caption{MIAB-augmented scenario: the MIAB trajectory (red stars) gravitates toward UE~1, which retains MIAB association (circle markers) at all snapshots, while UEs~2, 3, 4, and~5 remain FIAB-associated (triangle markers).}
        \label{fig:usecase_miab}
    \end{subfigure}
    \caption{Representative single-FIAB use case ($F=1$, $N=5$, $N_s=1$, $T=10$) illustrating Stage~5 UE--cell association decisions overlaid on the OKG-interpolated SINR heatmap. UE positions are labeled $j.t$ (UE index $j$, snapshot $t$). The high-attenuation region boundary is shown as a white contour.}
    \label{fig:usecase}
\end{figure}

This subsection quantifies the capacity gains from joint MIAB placement and association optimization (Stage~5) when augmenting an FIAB-only deployment with a single MIAB. The gain metric is defined as:
\begin{equation}
  G^{(t)} =
    \frac{C_{N_s}^{\mathrm{MIAB+FIAB},(t)}
        - C_{N_s}^{\mathrm{FIAB},(t)}}
         {C_{N_s}^{\mathrm{FIAB},(t)}}
    \times 100\,\%,
  \label{eq:gain}
\end{equation}
where $C_{N_s}^{\mathrm{MIAB+FIAB},(t)}$ and $C_{N_s}^{\mathrm{FIAB},(t)}$ denote the best aggregate capacity of the $N_s$ UEs with and without MIAB support, respectively.
The evaluation spans three FIAB densities ($F \in \{1,2,4\}$), three special-team sizes ($N_s \in \{1,2,5\}$ out of $N=5$ UEs), and two GA modes: standard (GA0: MaxGenerations$=700$, StallGenLimit$=500$) and fast (GA1: MaxGenerations$=100$, StallGenLimit$=50$), both with population size $150$, elite count $10$, crossover fraction $0.8$, and function tolerance $10^{-6}$. GA terminates when the relative change in the best fitness value over the stall generation limit falls below $10^{-6}$, or when the maximum number of generations is reached. Results are collected over $T=10$ trajectory snapshots, repeated three times per configuration.
Fig.~\ref{fig:usecase} illustrates a representative run ($F=1$, $N_s=1$, $T=10$). In the FIAB-only baseline (Fig.~\ref{fig:usecase_fiab}), all UEs associate with the single FIAB, with UE~5 persistently located in a low-SINR region ($<10\,\text{dB}$). In the MIAB-augmented scenario (Fig.~\ref{fig:usecase_miab}), Stage~5 repositions the MIAB at each snapshot toward UE~1, reassigning it to the MIAB and relieving FIAB contention, while remaining within a moderate-SINR corridor that ensures a viable backhaul link.
\begin{figure*}[!t]
    \centering    \includegraphics[width=\textwidth]{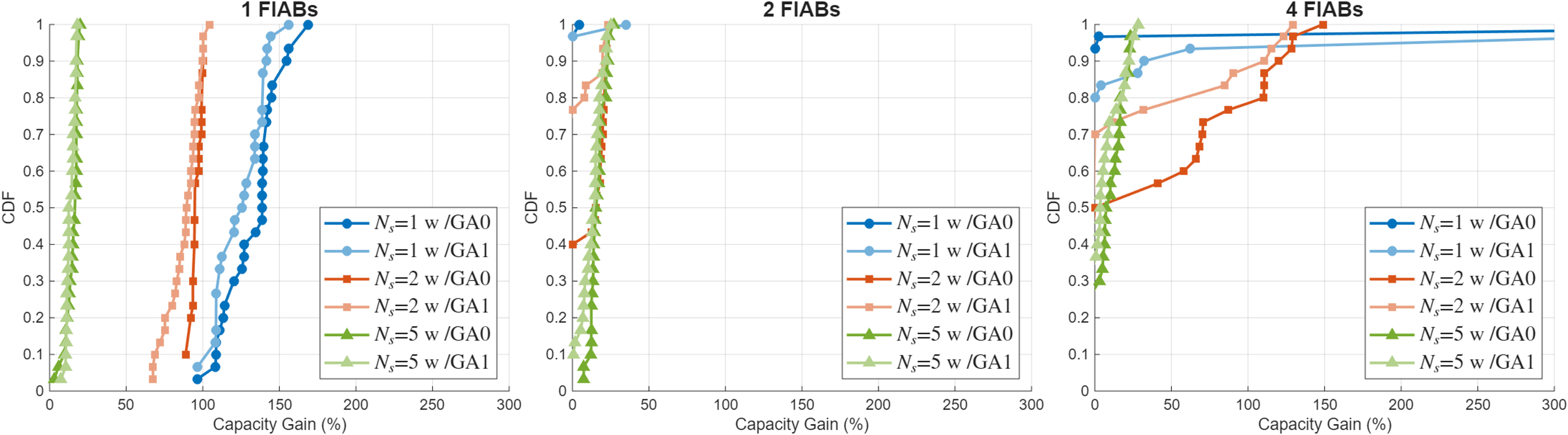}
    \caption{CDFs of the MIAB-augmented capacity gain $G^{(t)}$ defined in~\eqref{eq:gain} for three FIAB deployment scenarios ($F=1$, $F=2$, and $F=4$), three special-team sizes ($N_s \in \{1,2,5\}$), and two GA configurations.}
    \label{fig:cdf_all}
\end{figure*}
In the sparse deployment regime (left panel of Fig.~\ref{fig:cdf_all}), single-FIAB scenario ($F=1$), the MIAB acts as both a coverage enhancer and load offloader. For $N_s=1$ and $N_s=2$, median gains exceed $100\%$ and $\sim$$95\%$, with peaks up to $150\%$. For $N_s=5$, gains collapse below $30\%$ due to a backhaul bottleneck: the single FIAB must relay all MIAB-served traffic, limiting aggregate gain when resources are shared among all UEs.

With moderate infrastructure density (center panel), two-FIAB scenario ($F=2$), MIAB contribution becomes topology-dependent. For $N_s=1$, gains remain low ($<15\%$) as direct FIAB connectivity is already sufficient. For $N_s=2$ and $N_s=5$, gains increase to $10$--$40\%$ and $25$--$45\%$, respectively, reflecting scenarios where partial blockage allows the MIAB to improve both link quality and resource allocation.

In the dense deployment regime (right panel), four-FIAB ($F=4$), the gain distribution becomes highly heterogeneous. For $N_s=1$, gains are predominantly near $0\%$, except some extreme topology-driven edge cases. For $N_s=2$, medians exceed $50\%$ with upper tails reaching $150\%$, revealing an opportunistic regime where the MIAB selectively targets localized shadowing or load imbalances affecting UE subsets.
\begin{figure}[!t]
    \centering    \includegraphics[width=\columnwidth]{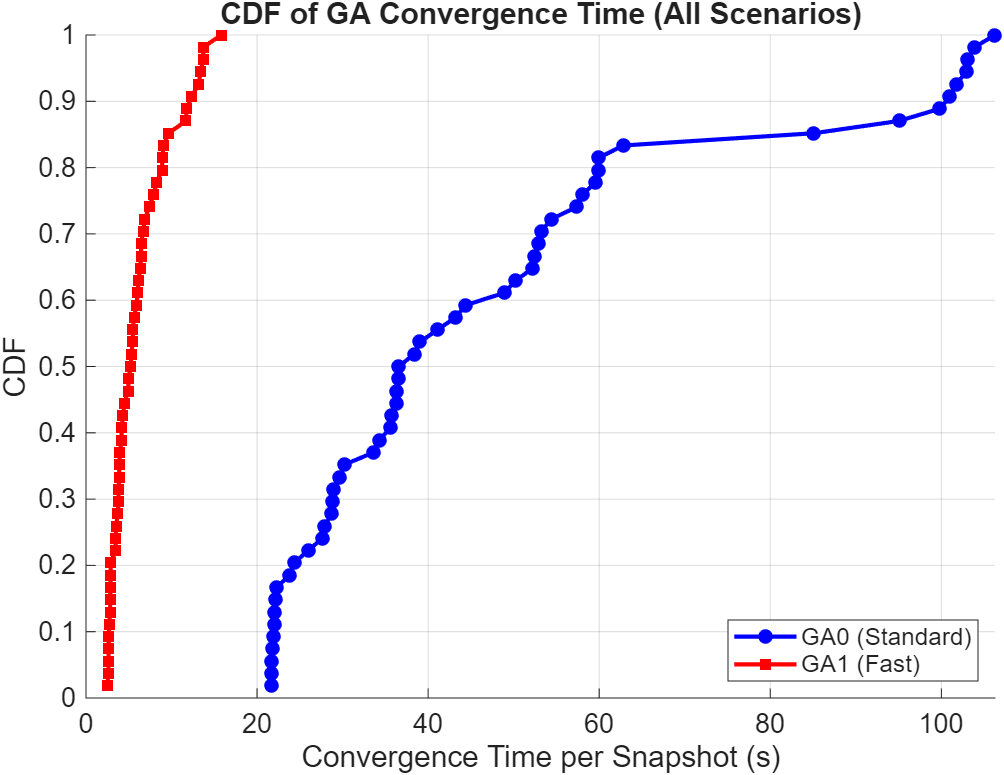}
    \caption{CDF of convergence time per snapshot for GA0 and GA1 across all evaluated scenarios.}
    \label{fig:CDFs_GA_time}
\end{figure}

Two main insights emerge. First, MIAB effectiveness is strongly dependent on baseline infrastructure density: in sparse deployments ($F=1$) it provides consistent coverage extension, while in dense deployments ($F=4$) its role becomes opportunistic. Second, $N_s$ introduces a trade-off between peak and aggregate performance: small $N_s$ enables highly targeted gains, while larger $N_s$ yields more stable but resource-constrained improvements due to shared backhaul limitations.

The fast GA configuration (GA1) closely tracks GA0 across all scenarios while converging within $5$--$15$~s per snapshot versus up to $100$~s for GA0 (Fig.~\ref{fig:CDFs_GA_time}), confirming near-optimal solutions at a fraction of the computational cost.

\subsection{PoC Field Demonstration}
\label{sec:field_validation}

To assess the practical applicability of the proposed pipeline, a PoC measurement campaign was conducted in a representative smart-port area measuring approximately $150\mathrm{m}\times130\mathrm{m}$. The demonstration area contained permanent cuboidal structures that were used as surrogates for the larger cuboidal blockage objects considered in the smart-port simulations and was discretized into geographical pixels of $5\mathrm{m}\times5\mathrm{m}$.

A total of $90$ measurement locations were selected across the area,
corresponding approximately to the $15\%$ sampling density identified as the
practical operating point in Sections~\ref{sec:results_kriging}
and~\ref{sec:results_clustering}. At each location, two measurements of RSRP and SINR were collected at UE and MIAB heights of $1.5\,\mathrm{m}$ and $2\,\mathrm{m}$ respectively, to emulate UE-level and MIAB-level observations.

\begin{figure}[t]
\centering
\includegraphics[width=\columnwidth]{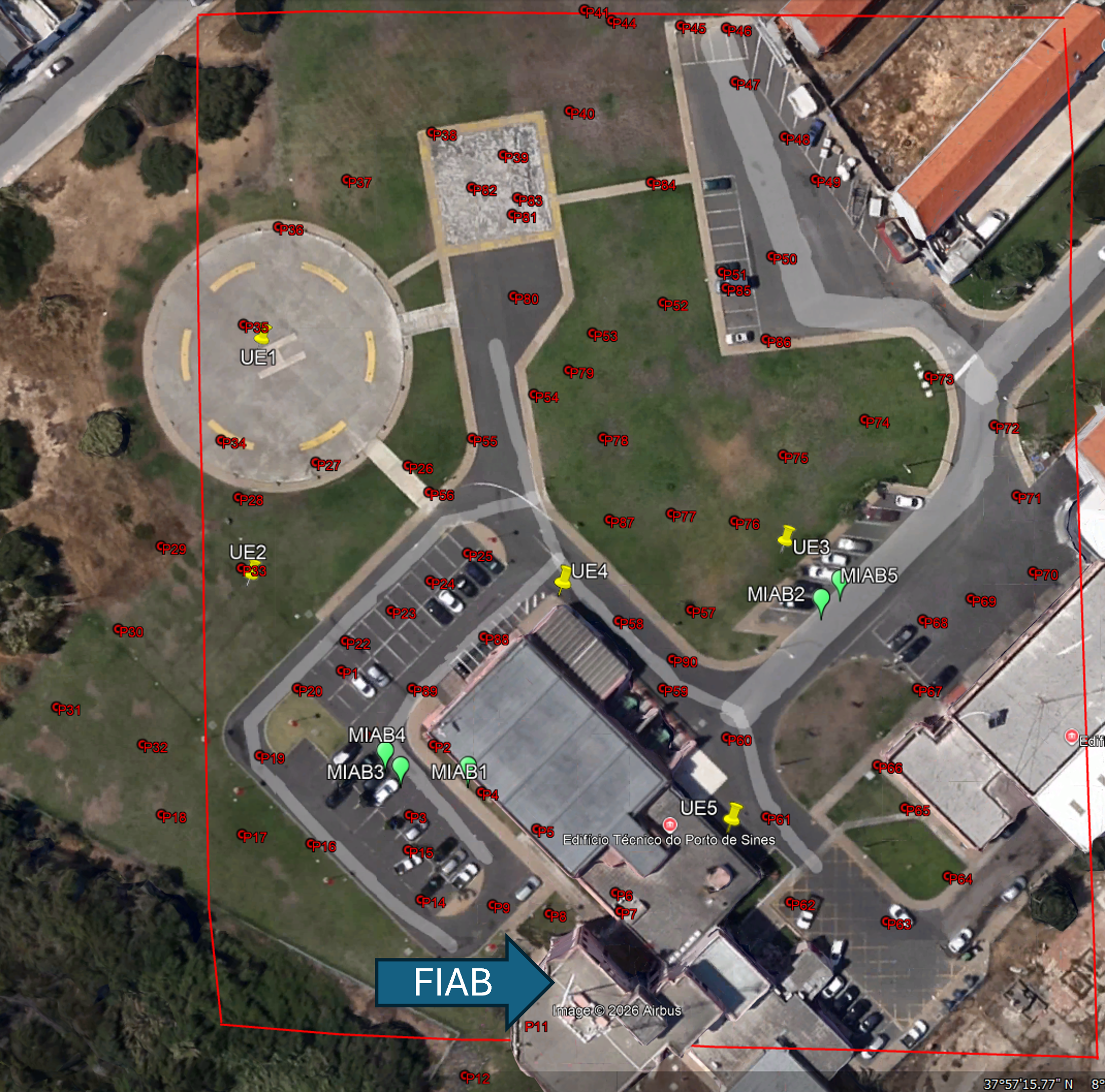}
\caption{Measurement area and sampled locations (red) used for the PoC field demonstration, showing the single FIAB location, the UE demonstration points (yellow pins), and the corresponding MIAB locations (green pins).}
\label{fig:field_campaign}
\end{figure}

\begin{figure*}[t]
\centering
\includegraphics[width=\textwidth]{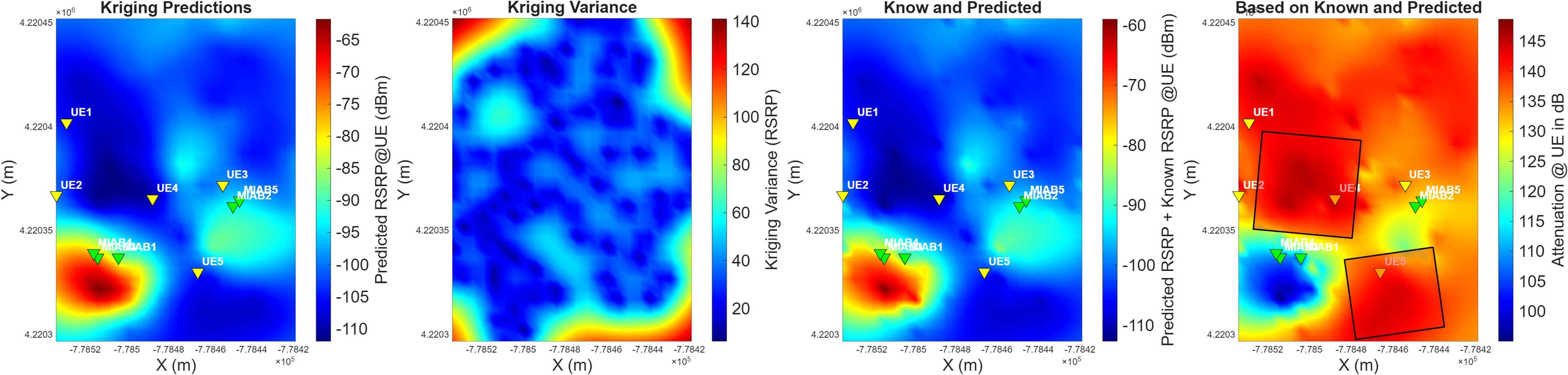}
\caption{Representative field-demonstration results for RSRP, including OKG prediction, prediction variance, reconstructed radio maps and attenuation with overwritten inferred cuboids; yellow and green pins indicate the UE demonstration points and corresponding MIAB locations, respectively.}
\label{fig:field_kriging}
\end{figure*}

\begin{figure}[t]
\centering
\includegraphics[width=\columnwidth]{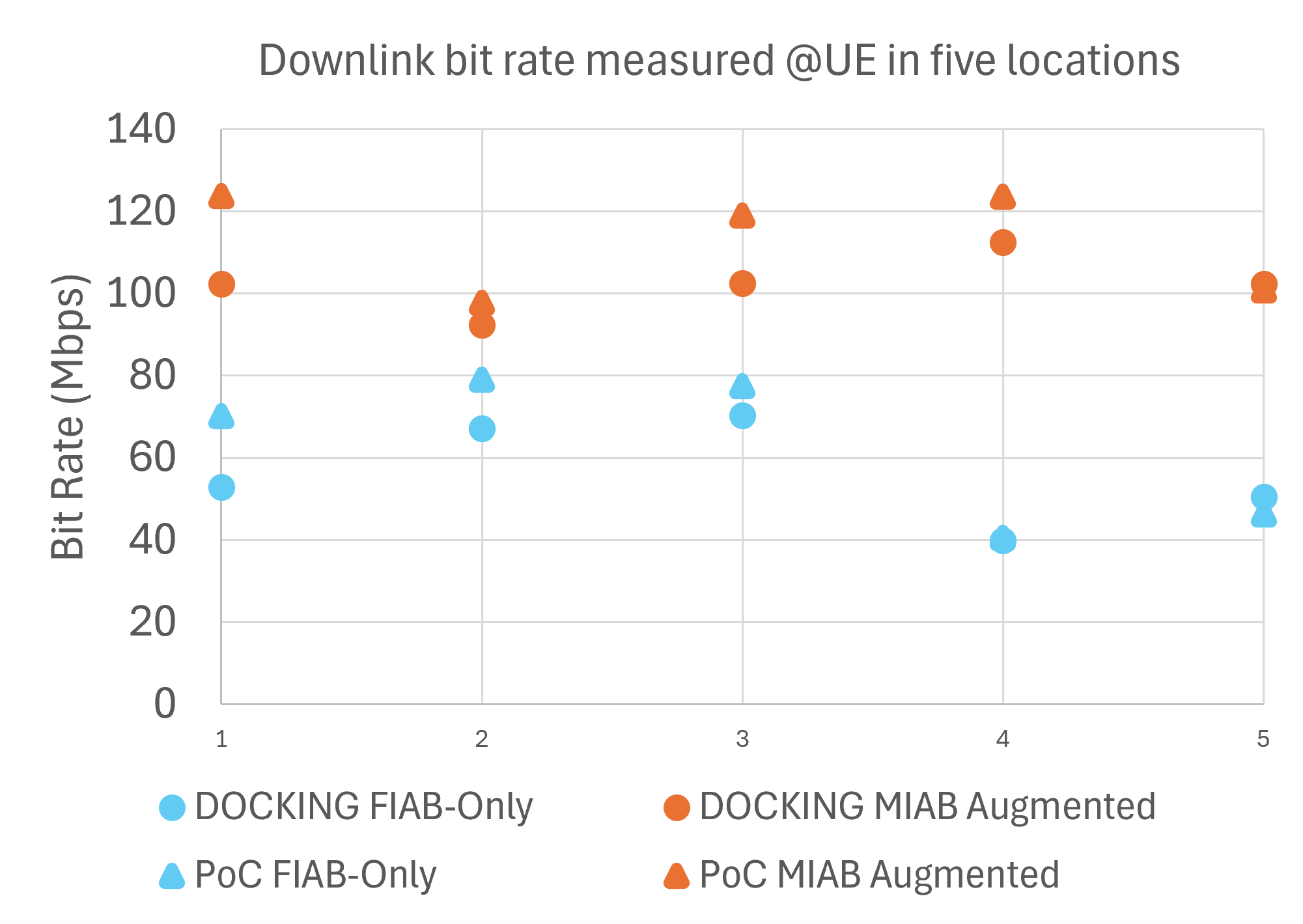}
\caption{Downlink UDP throughput at five PoC demonstration locations, comparing the FIAB-only baseline and MIAB-augmented configurations against Stage-5 predictions.}
\label{fig:field_throughput}
\end{figure}

Fig.~\ref{fig:field_campaign} illustrates the measurement area and sampling locations (red), including the UE demonstration points (yellow pins) and the corresponding MIAB locations (green pins). The collected measurements were processed through Stages~3 to~5 of the \textsc{DOCKING} pipeline; Stages~1 and~2 (ground-truth construction and sparse measurement emulation) were not applied, as real-world measurements replace the simulated dataset at this point. Fig.~\ref{fig:field_kriging} shows representative semivariogram fitting, OKG reconstruction, prediction variance, and reconstructed radio maps. Despite the sparse sampling density, the reconstructed RSRP, attenuation, and SINR distributions preserved the primary spatial propagation characteristics observed in the measurements.
The reconstructed attenuation map was subsequently processed by the obstacle-characterization stage. As illustrated in the rightmost column of Fig.~\ref{fig:field_kriging}, the
inferred cuboids were spatially aligned with the primary attenuation regions
associated with the permanent structures present in the area, confirming the ability of the proposed obstacle-characterization stage to identify optimization-relevant structures directly from sparse radio measurements. Finally, UDP downlink throughput measurements were collected using \texttt{iperf3} at five selected demonstration locations under both FIAB-only as baseline, and MIAB-augmented configurations. Fig.~\ref{fig:field_throughput} compares the measured throughput values against those predicted by stage~5. The predicted and measured
values follow consistent trends across locations, with the MIAB-augmented
configuration yielding higher throughput at all five points, confirming
that the capacity gain predicted by the optimization stage is realized in
practice. Although the PoC is limited in scale, these results provide experimental demonstration of the complete REM-to-obstacle-to-MIAB pipeline and confirm the practical applicability of Stages~3 through~5 of \textsc{DOCKING} beyond simulation-based evaluation.

\section{CONCLUSION}
\label{sec:conclusion}

This paper presented \textsc{DOCKING}, a unified REM-driven pipeline that transforms sparse radio measurements into obstacle-aware MIAB deployment decisions for smart-port environments, a framework to derive optimization-ready obstacle abstractions directly from reconstructed REMs, removing dependence on prior obstacle databases and explicit environment geometry. Using only sparse measurements, OKG reconstructed dense radio maps with prediction errors below $3\,\mathrm{dB}$ at the $90$th percentile using $15\%$ spatial sampling, while obstacle characterization achieved true-positive coverage above $85\%$ through compact cuboidal representations.

The inferred obstacle geometry enabled backhaul-aware MIAB placement, UE association, and backhaul selection. Capacity gains reached up to $150\%$ in sparse deployment scenarios, and the fast GA achieved near-equivalent performance while converging within $5$--$15\,\mathrm{s}$ per snapshot.

A PoC field campaign further corroborated the practical applicability of the proposed REM-to-obstacle-to-MIAB pipeline using real measurements, demonstrating that sparse radio observations, together with known network parameters and the considered modeling assumptions, can provide sufficient environmental awareness for deployment adaptation in obstruction-prone industrial environments.

Future work will extend from demonstration to validation toward larger operational seaport deployments and multi-MIAB scenarios.

\section*{ACKNOWLEDGMENT}
The authors would like to thank Pedro Duarte, Pedro Ribeiro and Hélder Fontes, from INESC TEC, for their support during the field PoC demonstration campaign conducted at the Port of Sines.

\begin{IEEEbiography}[{\includegraphics[width=1in,height=1.25in,clip,keepaspectratio]{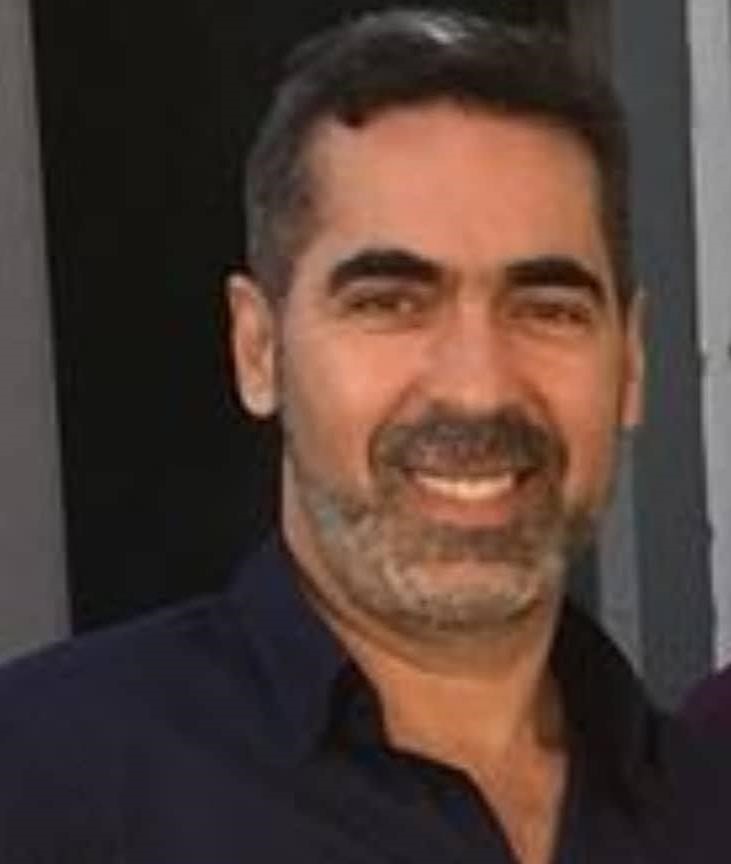}}]{PAULO FURTADO CORREIA } received the degree in Electrical and Computer Engineering from Instituto Superior Técnico (IST), Lisbon, Portugal, in 1991. He is currently pursuing his Ph.D.\ in Telecommunications at the University of Porto, conducting research with INESC TEC and participating in the NEXUS project. He is an Invited Assistant Professor at two private universities in Lisbon and has more than 20 years of professional experience across EMEA region in industry at vendors, operators, and consulting firms. His research interests include cellular technologies, IAB, O-RAN architectures, and performance modeling of 5G/6G networks. His work integrates extensive industry expertise with a growing academic portfolio, evidenced by several peer-reviewed publications and presentations at international conferences.
\end{IEEEbiography}

\begin{IEEEbiography}[{\includegraphics[width=1in,height=1.25in,clip,keepaspectratio]{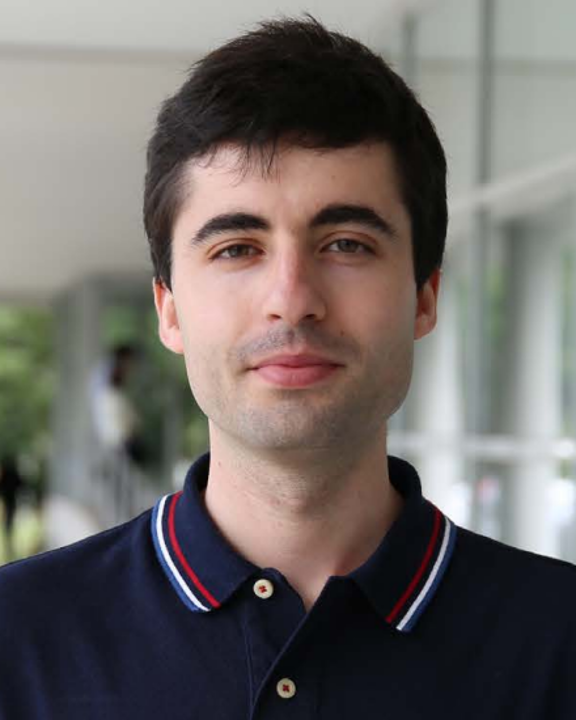}}]{ANDRÉ COELHO } (Member, IEEE) received his M.Sc. in Electrical and Computer Engineering in 2016 and his Ph.D. in Telecommunications in 2023, both from the University of Porto, Portugal. He is currently an Assistant Professor at the Faculty of Engineering of the University of Porto, where he teaches courses in telecommunications, and a Senior Researcher at INESC TEC, within the Centre for Telecommunications and Multimedia. He has authored 40+ peer-reviewed publications and participated in 15+ national and European research projects. His research interests include AI for wireless networks, O-RAN architectures, and autonomous network control and management, with a focus on on-demand network deployment using autonomous platforms. 
\end{IEEEbiography}

\begin{IEEEbiography}[{\includegraphics[width=1in,height=1.25in,clip,keepaspectratio]{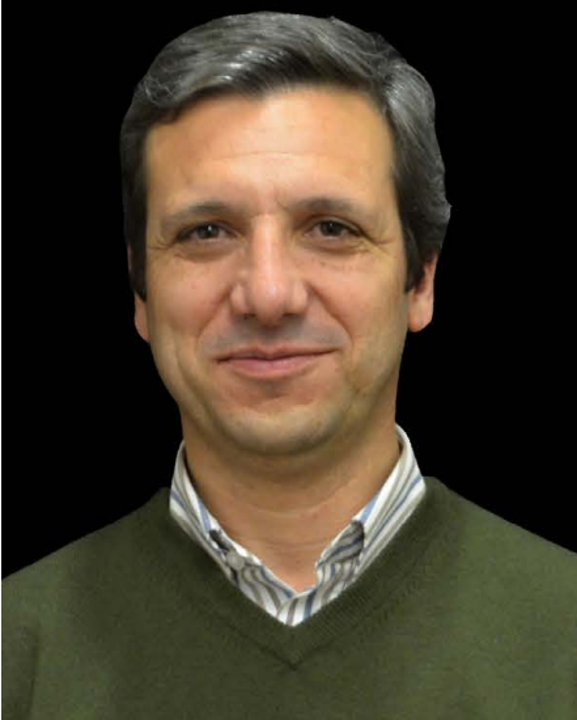}}]{MANUEL RICARDO } (Member, IEEE) received the Licenciatura, M.Sc., and Ph.D. (2000) degrees in Electrical and Computer Engineering (ECE), major of Telecommunications, from the Faculty of Engineering of the University of Porto (FEUP). He is a Full Professor at FEUP where he teaches courses on mobile communications and computer networks. He is the Director of the ECE Department at FEUP and a Director of the INESC TEC research institute. He participated in  40+ research projects and has more than 150 articles published. His research interests include wireless networks, quality of service, radio resource management, network congestion control, traffic engineering, and performance assessment.
\end{IEEEbiography}

\end{document}